\documentclass[aps,pra,amsmath,amssymb,11pt,final,tightenlines,twoside,onecolumn,nofloats,nofootinbib,superscriptaddress,
showkeys,showkeywords]{revtex4-2}

\usepackage[T2A]{fontenc}
\usepackage[utf8x]{inputenc}
\usepackage[russian,english]{babel}
\usepackage{graphicx}
\usepackage{longtable}
\usepackage[table]{xcolor} 

\newcommand{\meth}{~CH$_3$OH}
\newcommand{\ethyl}{~CH$_3$CH$_2$OH}
\newcommand{\acet}{~CH$_3$CHO}
\newcommand{\metf}{~CH$_3$OCHO}

\begin{document}

\noindent {\it Astronomy Reports , 2025, vol. , }
\bigskip\bigskip  \hrule\smallskip\hrule
\vspace{35mm}


\title{Multi-frequency mapping of the S255IR region at a wavelength of 1~mm}

\author{\bf\copyright~2025 \firstname{E.~A.}~\surname{Mikheeva}}
\email{e.mikheeva@lebedev.ru}
\affiliation{Lebedev Physical Institute, Russian Academy of Sciences, Astrospace Center, Moscow, Russia}

\author{\bf\firstname{S.~V.}~\surname{Kalenskii}}
\email{kalensky@asc.rssi.ru}
\affiliation{Lebedev Physical Institute, Russian Academy of Sciences, Astrospace Center, Moscow, Russia}

\author{\bf\firstname{S.-Y.}~\surname{Liu}}
\affiliation{Institute of Astronomy and Astrophysics, Academia Sinica, Taipei, Taiwan}

\author{\bf\firstname{A.~M.}~\surname{Sobolev}}
\affiliation{Xinjiang Astronomical Observatory, Chinese Academy of Sciences, Urumqi, China}
\affiliation{Ural Federal University, Yekaterinburg, Russia}

\author{\bf\firstname{S.}~\surname{Kurtz}}
\affiliation{Instituto de Radioastronomía y Astrofísica, Universidad Nacional Autónoma de México, Morelia, Mexico}

\begin{abstract}
\vspace{3mm}
\received{12.06.24}
\revised{12.06.24}
\accepted{12.06.24} 
\vspace{3mm}

The results of interferometric observations of the star-forming region S255IR in the frequency range 210--250 GHz are presented. The observations were carried out with the antenna array SMA (Hawaii, USA). We present the spectra of molecular cores SMA1 and SMA2 and the maps of the region for a large number of various molecular lines. In total, fifty-three molecules were detected, including complex organic molecules (COMs) such as CH$_3$CHO, CH$_3$CN, CH$_3$CH$_2$CN, and many others. Typical rotational temperatures in the hot core SMA1 fall in the range 100--200 K. Optical depths in the lines of methanol and some other molecules in the cores SMA1 and SMA2 were estimated. In SMA1, the optical depth of one of the strongest methanol lines, $5_{-1}-4_{-1}E$, proved to be $23.8 \pm 1.5$. Based on this value, one  can assume that the lines of other oxygen-containing COMs, such as CH$_3$OCHO, CH$_3$OCH$_3$, CH$_3$CH$_2$OH, which are typically much less abundant in hot cores than methanol, are optically thin in SMA1.

Most of the detected molecules can be roughly divided into two groups. The molecules of the first group emit exclusively toward the hot core SMA1, while some or all lines of the molecules of the second group, in addition to SMA1, can be seen toward a ring-like structure to the west of SMA1. This structure is most likely associated with the walls of a cavity formed by high-velocity outflows driven by young stellar objects (YSOs) in molecular cores SMA1, SMA2, and possibly SMA3. The gas temperature and density in the cavity walls were estimated using methanol lines. The temperature was found to be about 50--60 K, and the density about $10^7-10^8$ cm$^{-3}$. The column density of methanol near the brightness peaks in the lines of this molecule is about $5\times 10^{15}$~cm$^{-2}$. The column densities of other COMs in the ring-like structure will be determined in future studies with increased sensitivity achieved by spectral line stacking.
\end{abstract}

\maketitle

\section{INTRODUCTION}

The subject of this study is the high-mass star-forming region S255IR, located at a distance of 1.78 kpc from the Sun (\cite{burns2016h2o}) in the direction near the Galactic anticenter (l=192$^\circ$). It is the middle region in the chain of star-forming regions S255N, S255IR, and S255S. This chain is hosted by the same molecular cloud, sandwiched between the HII regions S255 and S257, which are part of the S254--S258 complex of ionized hydrogen regions. S255IR is the most evolved among the three regions in the chain; it contains three radio continuum sources~\cite{Snell_1986}; \cite{Ojha_2011}; \cite{Wang_2011} associated with H$_2$O masers, as well as with Class II CH$_3$OH masers~\cite{Minier_2000}; \cite{Goddi_2007}, which, in turn, indicate the presence of massive young stellar objects (MYSOs) in S255IR. Using MSX and IRAS data, Minier et al.~\cite{Minier_2005} estimated bolometric luminosity of S255IR (G192.60--MM2 in their notation). Assuming a distance of 2.6 kpc they obtained L$_{bol}=5.1\times 10^4~L_{\odot}$, which yields L$_{bol}$ recalculated for a distance of 1.78~kpc approximately equal to $2.4\times 10^4~L_{\odot}$.

\begin{figure}[h]
\caption{Map of the S255IR region in the  $5_{0}-4_{0}A^{+}$ methanol line at a frequency of 241791.367 MHz. Triangular markers indicate the dense cores SMA1 (coincident with NIRS3), SMA2, and SMA3 (coincident with NIRS1)}.
\label{fig:map}
\includegraphics[width=\textwidth]{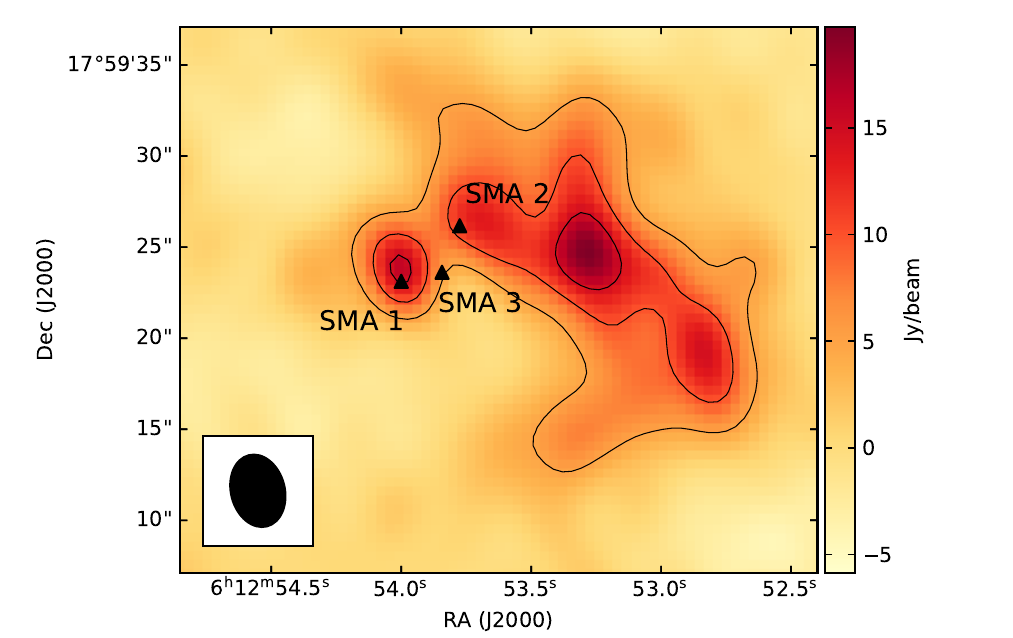}
\end{figure}

The region contains numerous infrared sources, the most prominent of which are NIRS1, whose nature is not entirely clear \cite{Simpson_2009}; \cite{Zinchenko_2020}, and a young stellar object (YSO) NIRS3 with a mass of about 20 M$_\odot$ and a bolometric luminosity of $\sim (2-4)\times 10^4~L_\odot$ (\cite{Wang_2011}; \cite{Zinchenko_2012}, \cite{Zinchenko_2015}). These objects coincide with the dense cores SMA3 and SMA1, respectively (e.g., \cite{Zinchenko_2020})(see Fig. \ref{fig:map}). In the mid-infrared range, NIRS3 appears as a disk seen almost edge-on (\cite{Boley_2013}). The axis of the large-scale ($\sim 1'$, which corresponds to a projected distance of $\sim 0.5$~pc) bipolar outflow, observed in CO and other molecular lines, is almost perpendicular to the disk (\cite{Zinchenko_2015}; \cite{wang2019elstag}).

Particular interest in the S255IR region arose after a Class II methanol maser (MMII) flare at 6.7 GHz was detected in 2015 towards the NIRS3 source (\cite{Fujisawa2015AFO}). Since MMII pumping is achieved by infrared radiation, immediately after the detection of this flare, observations of S255IR were carried out in the H (1.65 $\mu$m) and Ks (2.16 $\mu$m) bands, which showed an increase in brightness of $\Delta$H $\sim$3.5 magnitudes and $\Delta$Ks $\sim$2.5 magnitudes (\cite{Stecklum_2016}; \cite{Caratti_o_Garatti_2017}). Furthermore, the brightness of the infrared radiation toward the cavities, which were formed by the bipolar outflow from the protostar and scatter its radiation, increased noticeably. These facts indicate an outburst due to episodic accretion. Observations of the light echo propagation from the outburst between November 2015 and February 2016 established that the outburst began around mid-June 2015 (\cite{Caratti_o_Garatti_2017}). The results of near-infrared monitoring of the outburst are presented in \cite{Uchiyama_2019}. The outburst reached its maximum intensity in December 2015, after which a gradual decline began. By March 2019, the infrared emission intensity had returned to its pre-outburst level.

Various authors (\cite{burns2016h2o}, \cite{Wang_2011}; \cite{Zinchenko_2015}), based on observational data of water vapor masers and shock waves, have reported at least three outbursts preceding the 2015 event, with estimated ages of $\sim 7000$ years, $\sim 1000$ years, and $\le 130$ years. Apparently, outbursts in S255IR SMA1 occur episodically, similar to accretion outbursts in low-mass protostars. In the case of massive primordial star formation, episodic accretion bursts may play an important role in regulating the ionizing radiation emitted from the embedded massive star and providing a mechanism of prolonging accretion into the UCHII phase of massive star birth~(see~\cite{burns2016h2o} and references threin).

The next most intense source of submillimeter continuum in this region is the SMA2 core (\cite{Wang_2011}, \cite{Zinchenko_2012}), located approximately $4''$ northwest of SMA1. This core is also observed in lines of ammonia, methanol, and other molecules, and is likely the center of a bipolar outflow (\cite{Zinchenko_2020}, \cite{Zinchenko_2024}). 

The SMA3 core is associated with the NIRS1 source. As a result of near-infrared polarimetric observations, Simpson et al. (\cite{Simpson_2009}) discovered an S-shaped curved outflow oriented north-south and suggested that NIRS1 is a part of a binary system. However, Zinchenko et al. \cite{Zinchenko_2020} were unable to detect this outflow in the 3--2 CO line, possibly because the outflow lies close to the plane of the sky.

The 2015 outburst attracted significant research attention to the S255IR region. Numerous observations of the area were conducted across various spectral ranges, both in continuum and in lines of different molecules (\cite{Zinchenko_2012}, \cite{Zinchenko_2015}, \cite{Zinchenko_2020}, \cite{Zinchenko_2024}, \cite{Liu_2020}, \cite{Fedriani_2023} etc). Primarily, lines of simple molecules---CO, CS, SiO, CCH, NH$_3$, etc.---were observed and analyzed. Using data from the observations of these molecular lines, it was possible to estimate the main parameters of the SMA1 and SMA2 cores. It was found that SMA1 is a hot core with a temperature of about 170~K (\cite{Zinchenko_2015}), while for the SMA2 core the same study derived a temperature of 50 K. Liu et al. (2018) \cite{Liu_2018} detected lines of several complex organic molecules (COMs)---CH$_3$CHO, CH$_3$OCHO, CH$_3$CH$_2$OH, and others---in the spectrum of SMA1, but their analysis was limited to the CH$_3$CN lines.

Thus, relatively little is known about complex molecules in the S255IR region. Therefore, in 2020, we conducted observations of the S255IR region with the SMA interferometric array covering a wide frequency band of 210--250 GHz. This resulted in a dataset suitable for studying a large number of molecules. The aim of this work is to study the spatial distribution of various molecules---primarily COMs---in the S255IR region, as well as to conduct a comprehensive investigation of the molecular composition of various objects in this area. This paper presents maps of the emission from COM lines, as well as from lines of some simple molecules\footnote{A map in a line of a simple molecule is provided if no maps for this molecule's lines could be found in the literature.}. Furthermore, beam-averaged column densities and rotational temperatures of the molecules, derived using analytical methods assuming local thermodynamic equilibrium (LTE) in the source, are provided. More complete results of the spectral surveys, including non-LTE modeling results, will be presented in a separate paper.

\section{Observations and data calibration}
\label{sec:obs}
The S255IR region was observed on October 6, 2020, using the SMA Interferometric Array, consisting of eight six-meter telescopes located on the mountain Mauna Kea, Hawaii. The array was configured in a compact configuration, providing a spatial resolution of approximately $4''$ in the frequency range in which the observations were conducted. The 230 GHz (RxA) and 240 GHz (RxB) receivers were used, each receiving a single polarization plane; the planes are mutually orthogonal. 

The backend of each receiver was a correlator in 12 GHz Dual Rx mode, providing two 12-GHz bands separated by an 8-GHz window. As a result, the RxA receiver covered the 210–222 and 231–243 GHz frequency bands, and the RxB, the 218–230 GHz and 238–250 GHz bands. Each band, in turn, was divided into six 2.28-GHz  spectral windows with overlapping edges. The spectrometer resolution was 140 kHz, but it was smoothed during the data reduction to 1.117 MHz, which corresponds to a radial velocity resolution of 1.4 km/s at 240 GHz.

After observations the resulting array of visibility functions was converted at the radio telescope's Data Center to the Measurement Set format. This format is used for data reduction with the CASA software package, designed for processing radio interferometric data. The conversion was performed using the pyuvdata package~\cite{Hazelton2017}, \cite{Keating2025}; just at this stage the data were smoothed to a resolution of 1.117 MHz. Calibration and imaging were performed at the Astro Space Center of the Lebedev Physical Institute using the CASA package. First, the visibility functions were calibrated according to the SMA2 antenna array calibration manual\break (\verb|https://lweb.cfa.harvard.edu/rtdc/SMAdata/process/tutorials/sma_in_casa_tutorial.html|). Bandpass calibration was performed using the quasar 3C84. Point sources $0423-013$ and $0510+180$ were used as phase calibrators. Flux calibration was carried out using the solar system objects Vesta and Uranus. 

Using the CASA task {\em split}, the array of calibrated visibility functions was split into 24 parts, each corresponding to one spectral window. From each spectral window, the continuum was subtracted using the task {\em uvcontsub}. As a result, we obtained a dataset of calibrated continuum-free visibilities, which were further used to construct maps of S255IR in various molecular lines and spectra of this region in different directions.

\subsection{Map Construction Procedure}
\label{sec:maps}
The map construction process was as follows. First, using the task {\em tclean} in non-iterative mode (niter=0), a spectral cube of 2048 channels was constructed for each spectral window, where each channel represents an image of the source at a specific frequency, convolved with the array beam (\guillemotleft dirty\guillemotright map). For deconvolution, a square patch of sky centered at R.A.= 6$^h$12$^m$53.800$^s$, DEC = +17$^{\circ}59'22.09''$ (J2000) and 60 arcseconds in size was clipped from the spectral cube. This region includes NIRS1, NIRS3, and their environments. Subsequent cleaning was done in two stages. First, an initial, relatively rough deconvolution (cleaning of maps) was performed. This cleaning was performed using task {\em tclean} in non-interactive mode using a mask of dimensions $22''\times13''$ in right ascension and declination and was stopped when the standard deviation of the noise in the residual images stabilized. For quick viewing of the residual images, the CARTA\footnote{Cube Analysis and Rendering Tool for Astronomy, https://cartavis.org/} \cite{angus_comrie_2018_3377984} software was used. Second, a spectrum for the total observed frequency range was constructed. During the construction of this spectrum, the radiation intensity in each channel was averaged over the region, centered in the direction $06^h12^m53.62^s, 17^\circ59'22.2''$ ($J2000$) and having dimensions $27''\times 23''$, respectively, in right ascension and declination. This region includes all sources visible at this stage; thus, the spectrum was a net spectrum of SMA1, SMA2, and their environments. On this spectrum, using the catalog of spectral lines discovered in space\footnote{NIST Recommended Rest Frequencies for Observed Interstellar Molecular Microwave Transitions by Frank J. Lovas; https://physics.nist.gov/cgi-bin/micro/table5/start.pl}. by F. Lovas, lines of 27 molecules were identified, and {\em preliminary} maps were constructed for them.

Based on these preliminary maps, spectral lines were selected for further mapping. A list of these lines is given in Table ~\ref{tab:lines}. Most of selected lines arise from the molecules for which no detection in S255IR have been published. To construct the images, the continuum-free visibility functions in narrow spectral ranges around the line frequencies were transformed into spectral subcubes $60''\times 60''$ in RA and DEC, which were interactively cleaned with natural weighting. Using these subcubes, both channel maps and line integrated intensity maps were constructed.  The pixel sizes are $0.5''\times 0.5''$. To eliminate edge effects, the maps were cropped at the edges to a size of $50''\times 50''$. 

To create the maps, Python programs were written using the \guillemotleft aplpy\guillemotright, \guillemotleft numpy\guillemotright, \guillemotleft matplotlib\guillemotright, and \guillemotleft astropy\guillemotright~libraries.

The clean beam size decreases with increasing frequency from $4.3''\times 3.3''$ at 209.4 GHz to $3.8''\times 2.9''$ at 249.7 GHz, respectively. 

\subsection{Construction of SMA1 and SMA2 Spectra}
\label{sec:specs}
The spectra of SMA1 and SMA2 were constructed as follows. A region of $15''\times 12''$ around SMA1 was selected for cleaning, which contains the strongest sidelobes of the synthesized beam, as well as the SMA2 and SMA3 cores. In each of the 24 continuum-free spectral windows, this region was cleaned non-interactively, constructing spectral cubes. After 20,000 iterations, cleaning was stopped, and the spectrum of the region was examined. If the contribution of spectral lines above 3 sigma was no longer detectable in the spectrum constructed from the residual image cubes, cleaning was stopped. Otherwise, the next cleaning step was performed, and so on. In practice, no spectral window required more than three steps.

\begin{figure}
\caption{Spectrum toward SMA1}
\label{fig:spec_sma1}
\includegraphics[width=\textwidth]{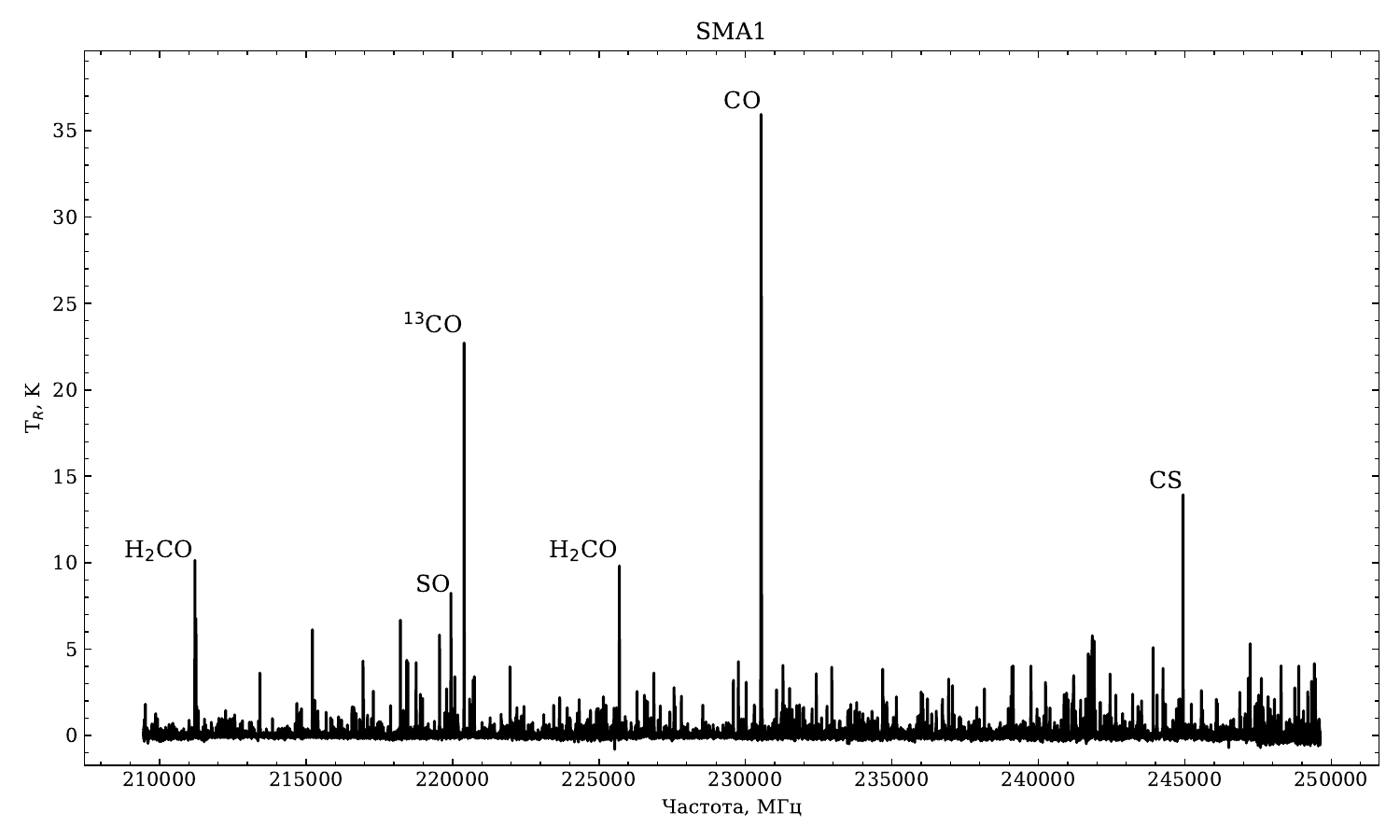}
\end{figure}

\begin{figure}
\caption{Spectrum toward SMA2}
\label{fig:spec_sma2}
\includegraphics[width=\textwidth]{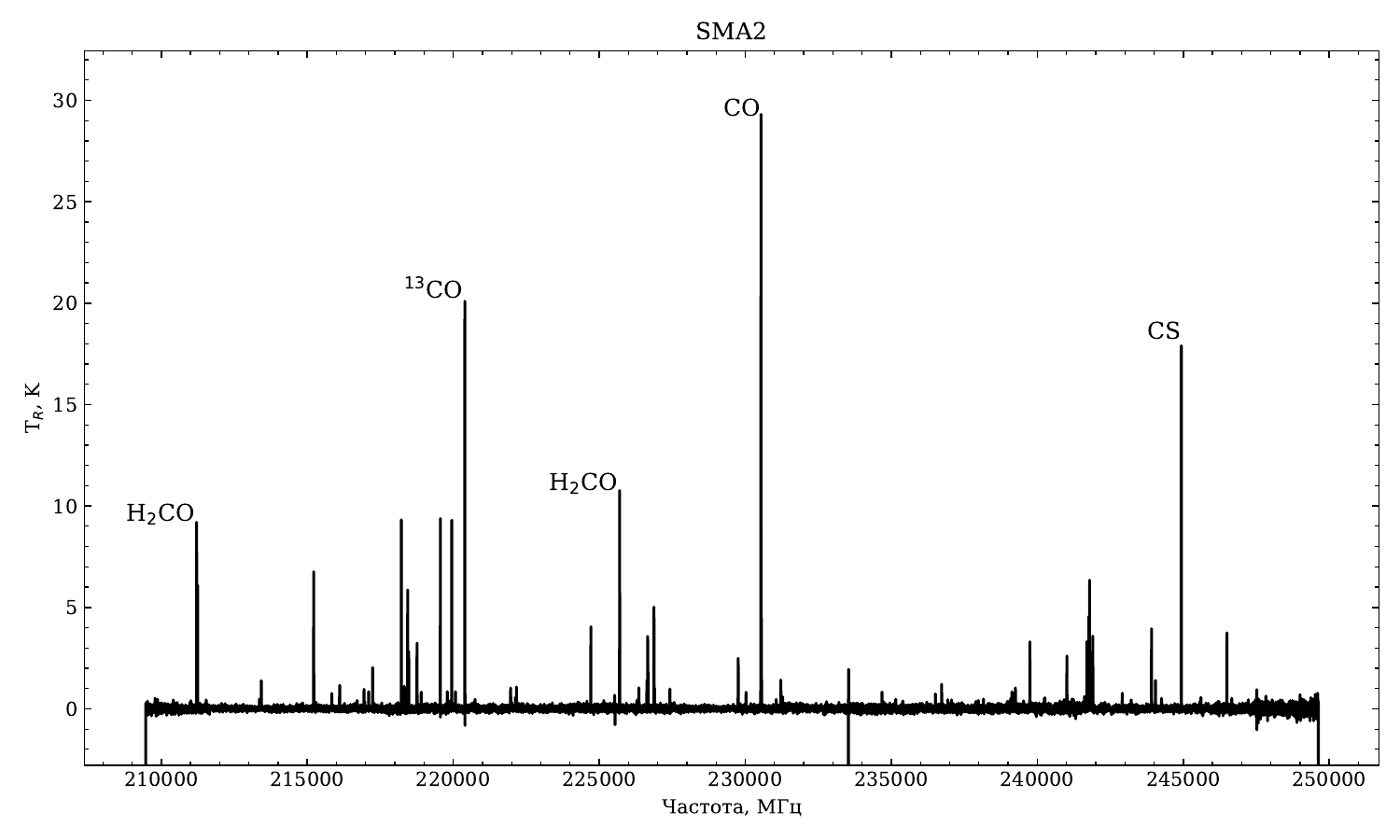}
\end{figure}

In the spectral cubes thus cleaned, regions centered on SMA1 and SMA2 and $4''\times 3''$ in size, which correspond to the angular sizes of the clean beam in declination and right ascension, were selected. Spectra were constructed for these regions, as shown in Figs.~\ref{fig:spec_sma1} and~\ref{fig:spec_sma2}.

 Numerous spectral lines were detected in the spectra of SMA1 and SMA2, the identification of which was carried out using the catalog of spectral lines discovered in space by F. Lovas (see section \ref{sec:maps}). However, many lines remained unidentified; their identification will be done in subsequent works during the analysis of the results of spectral surveys. 

\section{Maps of Integrated Line Intensities and Channel Maps}
\label{sec:maps_results}
The spectra of SMA1 and SMA2 contain numerous spectral lines belonging to various molecules, ranging from simple two- and three-atomic species--$^{13}$CO, C$^{18}$O, SiO, etc.--to complex organic molecules (COM)--\meth, \metf, \acet, \ethyl~etc. (Table~\ref{tab:found_molecules}). A list of the 53 detected molecules is given in table~\ref{tab:found_molecules}. Many of them have already been observed in S255IR, and their maps have been presented and analyzed in a number of papers (\cite{Zinchenko_2012}, \cite{Zinchenko_2015}, \cite{Zinchenko_2020}, \cite{Liu_2018}). With some exceptions (see below), we did not consider the lines of these molecules. However, molecules were also discovered that have not been mapped in S255IR previously. Among these are COMs \acet, \ethyl, CH$_3$CH$_2$CN, etc. We constructed maps in the lines of these molecules. In addition, the physical conditions necessary to produce emission in different transitions of the same molecule may be very different. Therefore, maps in different lines often do not coincide, and a set of maps allows us to characterize the source more completely than a map in a single line. An example is the maps of S255IR in two different methanol lines (Fig. ~\ref{fig:ch3oh}). Therefore, we also constructed maps of S255IR in the lines of methanol, methyl cyanide and some other molecules observed in this source earlier. The complete list of molecules for which maps were constructed in one or more lines is given in Table ~\ref{tab:found_molecules}, and the list of lines for which we constructed maps is presented in Table ~\ref{tab:lines}. The constructed maps are available online (http://www.asc.rssi.ru/kalenskii/s255ir/s255ir.html).

\begin{table}[htbp]
\centering
\caption{Molecules detected in the cores SMA1 and SMA2.}
\label{tab:found_molecules}
\begin{tabular}{|l|l|l|}
\hline
       & SMA1                                                 & SMA2 \\
\hline
Diatomic&СO, C$^{18}$O, C$^{17}$O, $^{13}$CO, CN, C$^{15}$N, CS, & CO, C$^{18}$O, C$^{17}$O, $^{13}$CO, CN, CS,\\
        & C$^{34}$S, $^{13}$CS, SiO, SO, $^{33}$SO, $^{34}$SO, S$^{18}$O & C$^{34}$S, $^{13}$CS, $^{34}$SO, SO, SiO\\
\hline
Triatomic& DCN, H$_2$S, HDO, OCS, OC$^{34}$S, SO$_2$,           & DCN, H$_2$S, OCS, SO$_2$\\
      & $^{34}$SO$_2$, HCS$^+$ & \\
\hline
Four-  & H$_2$CO, H$_2^{13}$CO, H$_2$CS, HNCO                & H$_2$CO, H$_2^{13}$CO, H$_2$CS, HNCO \\
atom   &          &\\
\hline
Five-  & CH$_2$CO, HCCCN, H$^{13}$CCCN, HC$^{13}$CCN,           & c-C$_3$H$_2$, HCCCN\\
atom   & HCOOH, HCOOD, c-C$_3$H$_2$?, CH$_2$NH                &  \\
       &  NH$_2$CN                                         &\\
\hline
Six-   & CH$_3$CN, $^{13}$CH$_3$CN, CH$_3^{13}$CN, CH$_3$OH, CH$_3$OD &  CH$_3$OH, $^{13}$CH$_3$OH, CH$_3$OD\\
atom   & $^{13}$CH$_3$OH, CH$_3$SH?, NH$_2$CHO    &\\
\hline
Seven- & CH$_3$CHO, CH$_3$CCH, $^{13}$CH$_3$CCH, CH$_3$NH$_2$ & CH$_3$CCH\\ 
atom   & CH$_2$CHCN, c-C$_2$H$_4$O                         & \\
\hline
Eight- & CH$_3$OCHO                                            & \\
atom   & &\\   
\hline
Nine-  & CH$_3$CH$_2$CN, g-CH$_3$CH$_2$OH, t-CH$_3$CH$_2$OH,  &\\
atom   & CH$_3$OCH$_3$&\\ 

\hline
\end{tabular}
\end{table}

\section{Results}
\label{sec:results}
\begin{figure}
\caption{Maps in the $12_{0,12}-11_{0,11}E$ ethanal (CH$_3$CHO) line, in the $22_{0,22}-21_{0,21} \quad A$ methyl formate (CH$_3$OCHO) line, in the $6_{3,4}-5_{2,3}$ ethanol (t-CH$_3$CH$_2$OH) line, and in the $25_{4,21}-24_{4,20} $ propionitrile (CH$_3$CH$_2$CN) line.}
\label{fig:multimap}
\includegraphics[width=0.95\textwidth]{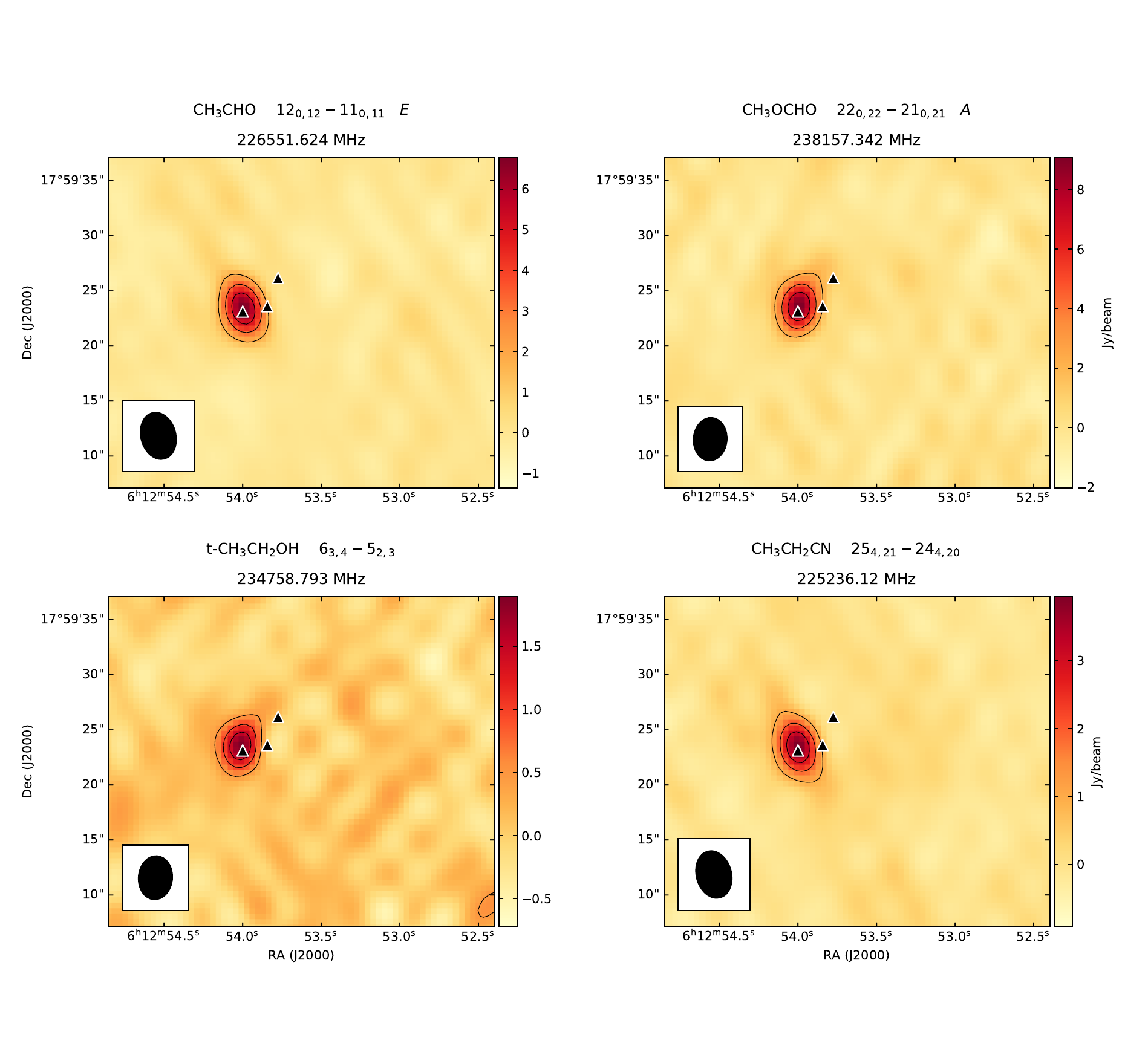}
\end{figure}

Based on the mapping results, all detected molecules can be divided into two large groups. The first group consists of HNCO, OCS, CH$_3$CHO, CH$_2$CO, CH$_3$OCHO, CH$_3$CN, and CH$_3$CH$_2$CN. In all lines of these molecules, emission is visible only toward SMA1 and the sources are not spatially resolved with our beam. As an example, Fig.~\ref{fig:multimap} shows the integrated intensity maps of the lines of ethanal (CH$_3$CHO), methyl formate (CH$_3$OCHO), ethyl alcohol (CH$_3$CH$_2$OH), and ethyl cyanide (CH$_3$CH$_2$CN). The remaining maps in the lines of these molecules demonstrate a similar source structure, differing only in intensity. 

\begin{figure}
\caption{Samples of spectral lines of the first group molecules. The figure shows the $11_{1,10}-10_{1,9}$ CH$_2$CO line, the $12_4-11_4$ CH$_3$CN line, the $12_{0,12}-11_{0,11}E$ CH$_3$CHO line, the $19_{3,17}-18_{2,16}E$ CH$_3$OCHO line, the $24_{3,22}-23_{3,21}$ CH$_3$CH$_2$CN line, the $10_{1,10}-9_{1,9}$ HNCO line, and the $18-17$ OCS line. Antenna temperature of the CH$_3$CH$_2$CN line is multiplied by six.}
\label{fig:smaspecs}
\includegraphics[width=1\textwidth]{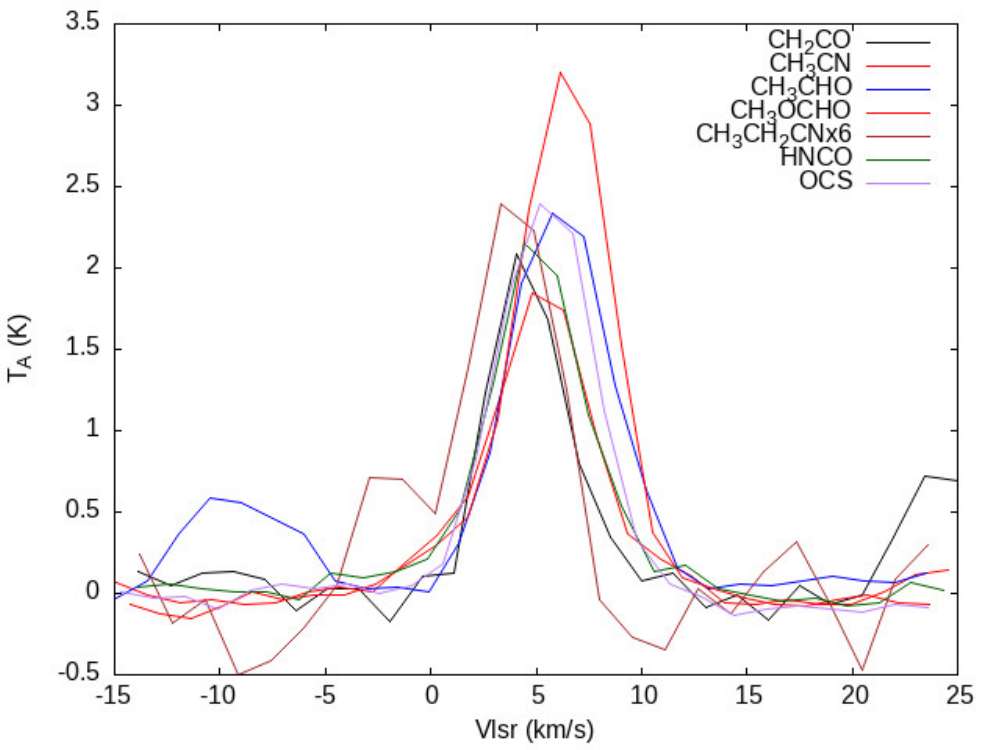}
\end{figure}

Figure~\ref{fig:smaspecs} demonstrates line profiles of the first group molecules. One can see that the LSR radial velocities are close to 5~km/s.

All or some lines of molecules of the second group emit not only in the SMA1 core, but also in other directions. Moreover, the emission of SMA1 is not necessarily the brightest, and in some cases is absent altogether. 

\begin{figure}
\caption{Upper row: maps of the integrated intensity of the $J=3/2-1/2~F=5/2-3/2$ CN line (left) and of the $3_{0, 3}-2_{0, 2}$ H$_2$CO line (right).
Lower row: maps of the 5--4 HCS$^+$ line (left) and of the 5--4 CS line (right).}
\label{fig:cn}
\includegraphics[width=\textwidth]{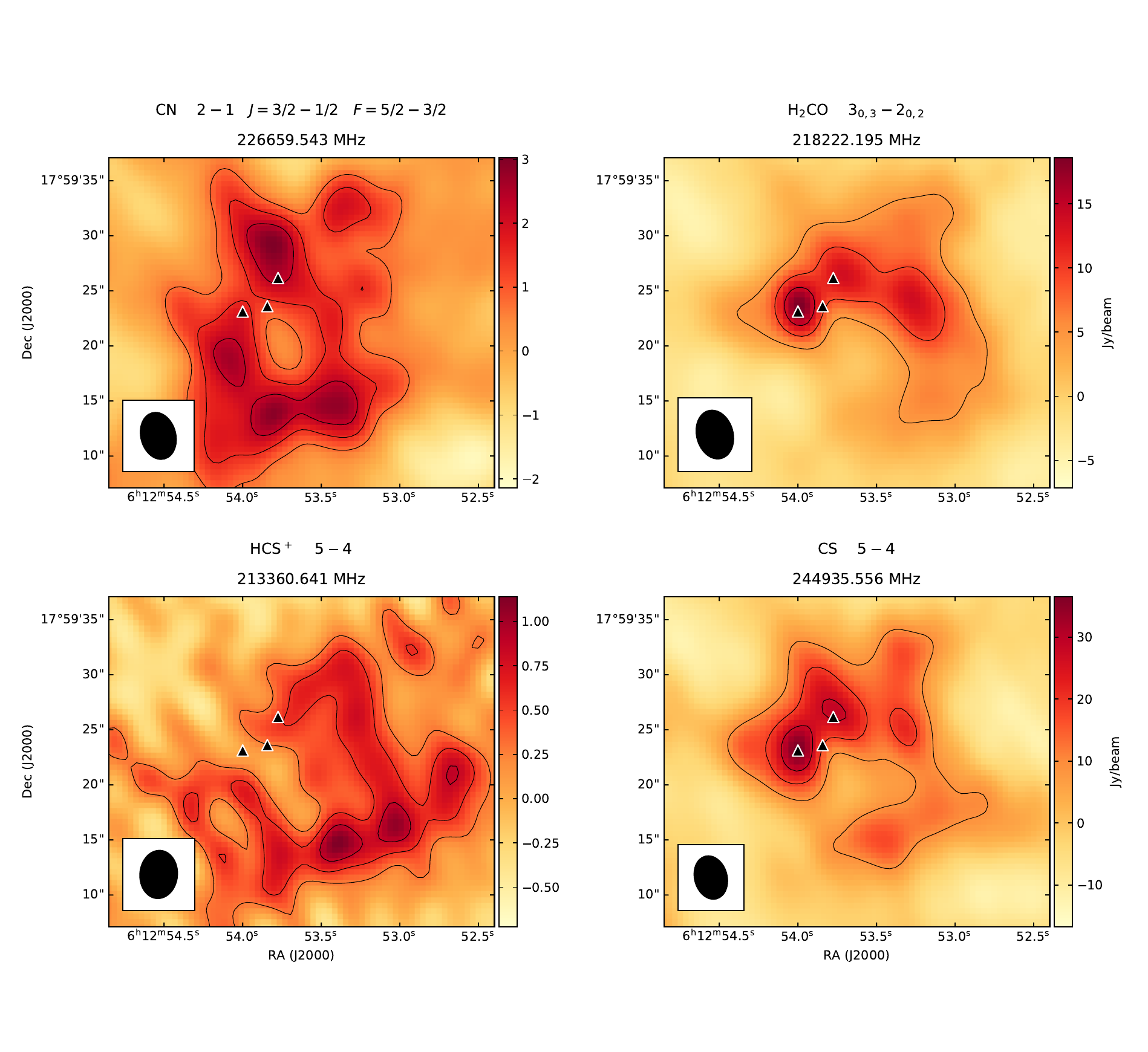}
\end{figure}

One of the molecules of the second group is cyanogen (CN). We constructed maps in five hyperfine components of the $2-1$ rotational transition. Fig.~\ref{fig:cn} shows the map of CN in the $J=3/2-1/2~F=5/2-3/2$ line. One can see that the source has a ring-shaped structure about $15''$ in diameter with several brightness peaks. One of the brightest peaks is located toward SMA2, while the emission brightness toward SMA1 is very low. The maps in the remaining detected $J=3/2-1/2$ lines generally coincide with the map in the $J=3/2-1/2~F=5/2-3/2$ line. All images show the same ring-shaped structure with the brightest peak toward the SMA2 core and the second brightest component  located approximately diametrically opposite SMA2. In contrast, the image of the spectral feature at $\approx 226875$~MHz, which is a blend of three hyperfine components of the $J=5/2-3/2$ transition is noticeably different. The brightest region at 226875~MHz is extended and roughly coincides with the continuum structure discussed by e.g. Zinchenko et al. (\cite{Zinchenko_2020}), and the peaks in the ring structure are two to three times less bright than this region.

The maps in the lines of some molecules resemble the CN ring with one or two breaks or a portion of this ring. For example, the image in the $5-4$ HCS$^+$ line at a frequency of 213360.650 MHz (Fig.~\ref{fig:cn}) resembles this ring, but with the most noticeable break toward SMA1 and SMA3. The brightest peak approximately coincides with the bright peak on the CN maps, which is located diametrically opposite to SMA2. Another such molecule is CS. The image in the 5--4 CS line at 244935.556 MHz (Fig.~\ref{fig:cn}) has a ring shape with the brightest peaks toward SMA1 and SMA2 and a break at a distance of about $7''$ to the southwest of SMA1.

Images of the lines of other molecules---SiO and H$_2$S---do not show a ring-shaped structure. However, the regions emitting in the lines of these molecules are often located within the CN ring. Therefore, it is likely that these regions and the ring observed in the lines of CN and other molecules belong to the same physical structure, which will be discussed below.

\begin{figure}
\caption{Maps in methanol lines $5_{2}-4_{2} A^+$ at 241887.674~MHz (left) and $4_{2}-3_{1} E$ at 218440.05~MHz (right)}
\label{fig:ch3oh}
\includegraphics[width=1\textwidth]{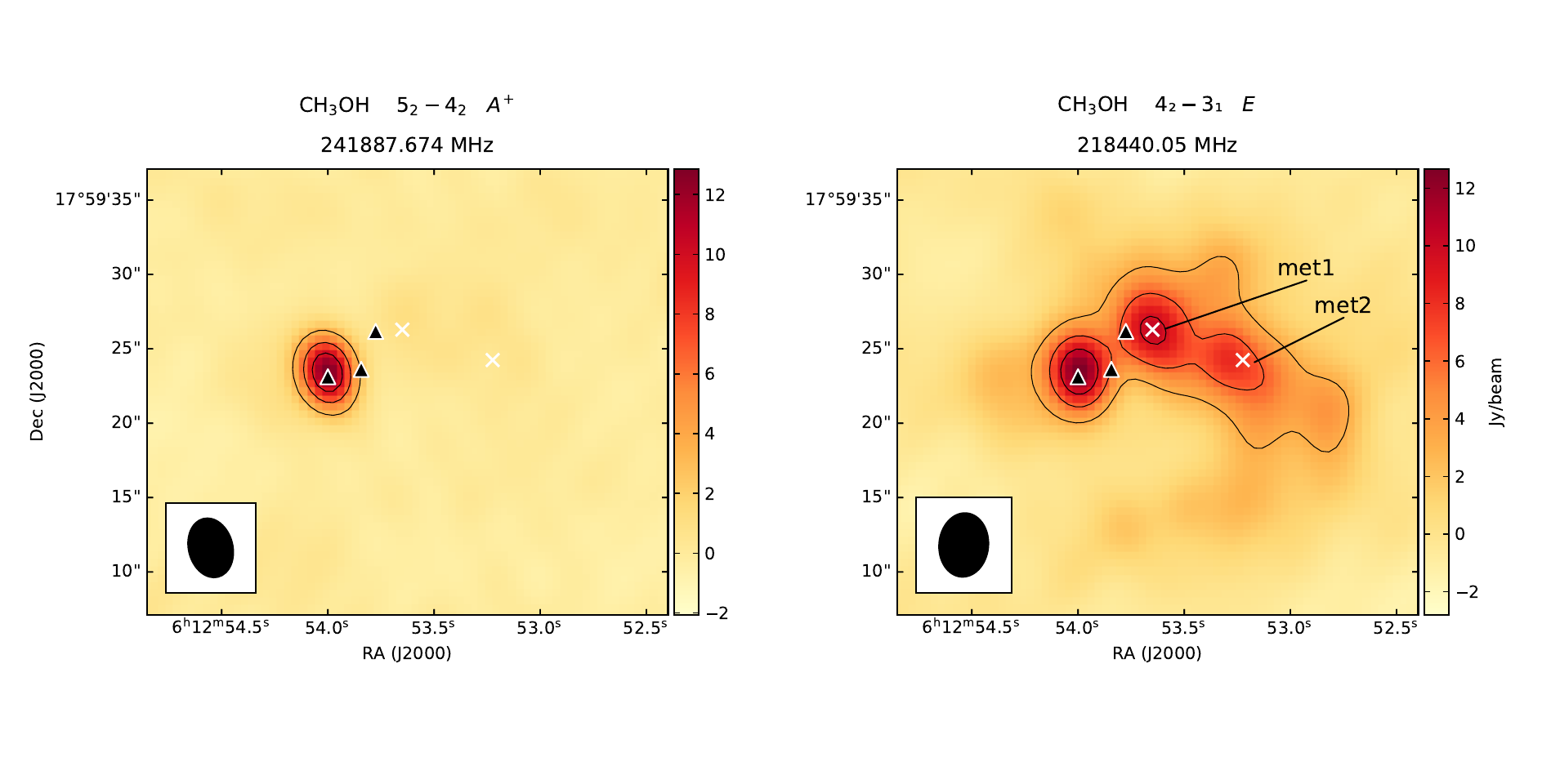}
\end{figure}

A large number of maps have been constructed for various methanol lines, two of which are shown in Fig. \ref{fig:ch3oh}. The emission in the lines with lower-level energy $E_{low} \gtrsim 50$~K is concentrated in the SMA1 direction (Fig. \ref{fig:ch3oh}, left), while lines with $E_{low} < 50$~K show extended emission, as do the other Group II molecules (Fig. \ref{fig:ch3oh}, right).

The low-level energies of all mapped lines of COMs CH$_2$CO, CH$_3$CHO, C$_2$H$_5$CN, CH$_3$CN, CH$_3$OCHO are above 50~K. It is possible that images of low-energy lines of these molecules would show a similar morphology to maps of molecules of the second group. It would be interesting to construct such maps and check whether these molecules are indeed present in gas phase only in the hot core SMA1. In addition, it is possible that the weak extended emission in the lines of at least some of the molecules of the first type (for example, H$_2^{13}$CO) is not visible due to insufficient sensitivity.
It's possible that some or even all of the molecules in the first group, like methanol, would show emission in regions other than SMA1 when observed with higher sensitivity. This issue will be investigated in future studies.

\section{Rotation diagrams}
\label{sec:rot}
\begin{figure}
\caption{Rotation diagrams built for the molecules detected toward SMA1}
\label{fig:rot_sma1}
\includegraphics[width=0.8\textwidth]{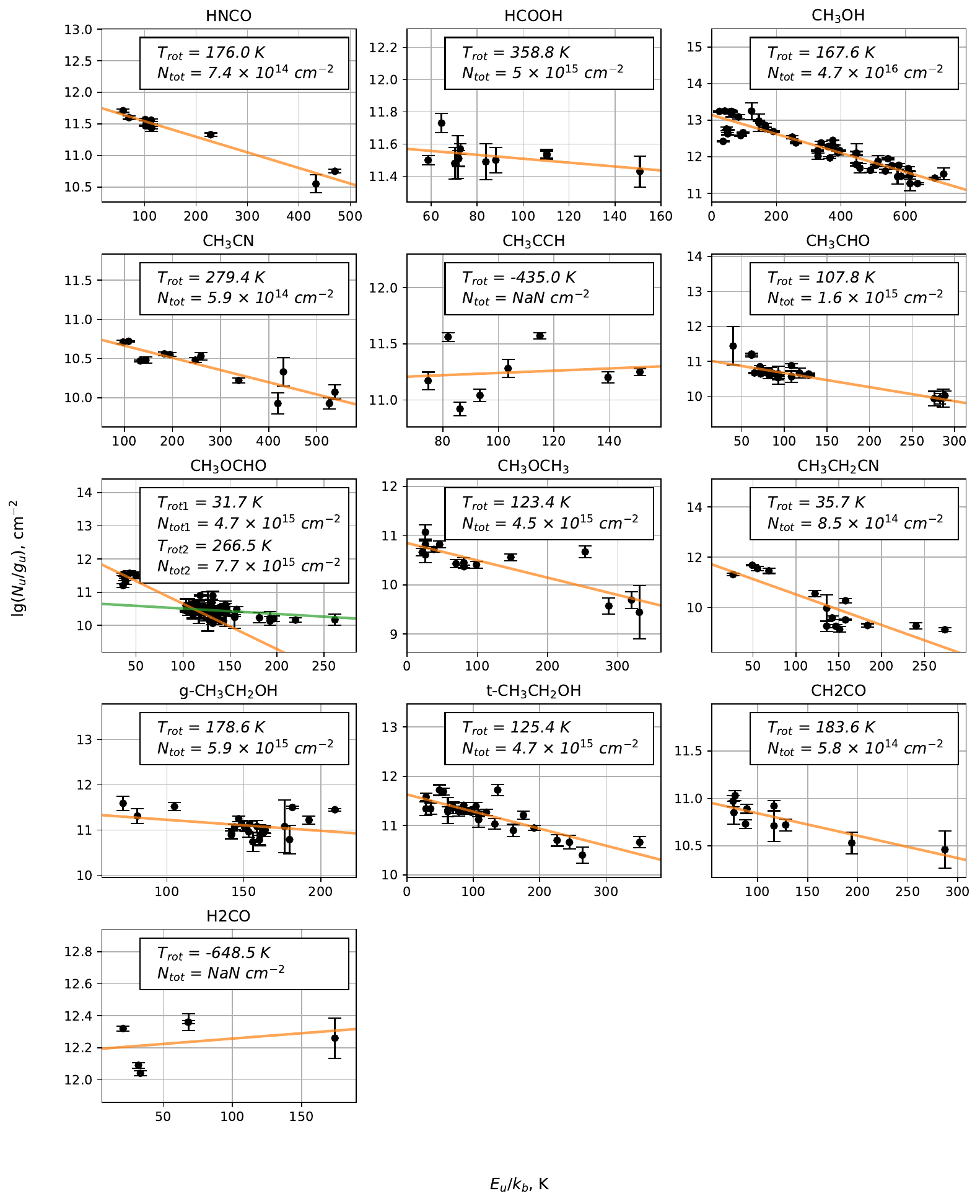}
\end{figure}

If four or more isolated lines of a molecule were detected in a single direction, we built a rotation diagram (RD) for this molecule. The RDs constructed for the molecules in the SMA1 direction are shown in Fig.~\ref{fig:rot_sma1}. The scatter of points on the diagrams is fairly large, which in turn leads to fairly large uncertainties of both the rotational temperatures and the column densities of the molecules. This scatter may be caused by deviations from LTE; as for the optical depth, we do not believe that it could significantly affect the RDs built for the most of the detected COMs (see sect.~\ref{sec:opt_depth}). In addition to deviations from LTE, incorrect identification of lines may play a certain role. 

Rotational temperatures ($T_{rot}$) for most molecules fall in the range 100--200~K. Higher $T_{rot}$ values were obtained for HCOOH ($T_{rot}$ = 359~K) and CH$_3$CN ($T_{rot}$=279~K). Much lower value ($T_{rot}=36$~K) was found for CH$_3$CH$_2$CN\footnote{This rotation diagram may be two-component, similar to the RD of methyl formate, but the number of points corresponding to $E_{up}\gtrsim 150$~K is not large enough to reliably establish the existence of a high-temperature component.}. The rotational temperature of formaldehyde, as well as of methylacetylene, which is considered a good thermometer of interstellar gas, was found to be negative. We believe that in the case of methylacetylene this may be caused by the absorption in the K=0 and 1 lines in a cold shell\footnote{The results of modeling the CH$_3$CCH emission will be presented in the following paper.}; in the case of formaldehyde, in addition to the hypothetical shell, the absence of LTE and the large optical depth of the lines could have influenced the RD shape. The rotation diagram of CH$_3$OCHO is two-component: in the range of level energies $0-100$~K, the rotational temperature is equal to 32~K\footnote{Due to the absence of points in the range of level energies 70--100~K, the rotational temperature of the cold component is poorly determined and the obtained value is an upper limit.}, and at level energies above 100~K its value is equal to 266~K.

\begin{table}[]
    \centering
    \begin{tabular}{lll}
    \hline
Molecule \qquad & Rotational temperature & \quad  Column density\\
                 & (K) & \quad (cm$^{-2}$) \\
        \hline
        H2CO & --649  15 & \quad NaN \\
        HNCO & 176 $\pm$ 15 & \quad (7.4 $\pm$ 0.8) $\times$ 10$^{14}$\\
        HCOOH & 359 $\pm$ 236 & \quad (5.0 $\pm$ 0.9) $\times$ 10$^{15}$\\
        CH$_2$CO & 184 $\pm$ 37 & \quad (6 $\pm$ 1) $\times$ 10$^{14}$\\
        CH$_3$OH & 168 $\pm$ 9 & \quad (4.7 $\pm$ 0.6) $\times$ 10$^{16}$\\
        CH$_3$CN & 279 $\pm$ 39 & \quad (5.9 $\pm$ 1.0) $\times$ 10$^{14}$\\
        CH$_3$CCH & -435 $\pm$ 1343 & \quad NaN \\
        CH$_3$CHO & 108 $\pm$ 12 & \quad (1.6 $\pm$ 0.2) $\times$ 10$^{15}$\\
        CH$_3$OCHO & 32 $\pm$ 4 & \quad (5 $\pm$ 2) $\times$ 10$^{15}$\\
        CH$_3$OCHO & 266 $\pm$ 165 & \quad (8 $\pm$ 3) $\times$ 10$^{15}$\\
        CH$_3$OCH$_3$ & 123 $\pm$ 21 & \quad (4.5 $\pm$ 1.2) $\times$ 10$^{15}$\\
        CH$_3$CH$_2$CN & 36 $\pm$ 6 & \quad (8.5 $\pm$ 9.1) $\times$ 10$^{14}$\\
        g-CH$_3$CH$_2$OH & 179 $\pm$ 112 & \quad (5.9 $\pm$ 4.3) $\times$ 10$^{15}$\\
        t-CH$_3$CH$_2$OH & 125 $\pm$ 17 & \quad (4.7 $\pm$ 0.8) $\times$ 10$^{15}$\\
        \hline
   \end{tabular}
    \caption{Rotational temperatures and column densities for the molecules detected toward SMA1}
    \label{tab: TandN_table}
\end{table}
Rotation diagrams were constructed for the hot core SMA1, as just here we found four or more lines for a number of molecules. In this object Zinchenko et al.~\cite{Zinchenko_2015} found a temperature of about 170~K. Liu et al~\cite{Liu_2020} modelled CH$_3$CN emission in SMA1 and found that the temperature is higher than 400~K in the inner regions, gradually falling off to the temperature of $\gtrsim 100$~K toward the peripheral areas. Most of the molecules for which RDs were constructed belong to the first group, meaning that their emission was detected only in SMA1; the exceptions are methanol, methyl acetylene, and formaldehyde. Therefore, it is not surprising that in most cases the rotational temperature exceeds 100~K. However, the rotational temperature of ethyl cyanide is only 36 K, and the rotational temperature of the cold component of methyl formate is below 32 K. These facts indicate that the hot core is accompanied by cold gas, and the apparent angular dimensions of the cold source are smaller than the synthesized beam.

It should be noted that the presence of cold gas in SMA1 has been noted in previous studies. For example, Zinchenko~et~al.~\cite{Zinchenko_2015} constructed a rotation diagram for SO$_2$ in SMA1 based on observations with a spatial resolution of $2''$, which also turned out to be two-component. For low-energy levels, the rotational temperature was found to be $65\pm 11$~K, and for high-energy levels, $146\pm 16$~K. Zinchenko et al. explained the presence of the cold component by the contribution of a cloud surrounding the hot core. 
Moreover, based on the detection of a strong DCN emission near the center of the hot core, Zinchenko et al.~\cite{Zinchenko_2015} suggested a presence of a relatively cold (<80 K) and rather massive clump there. 

DCN is usually tracing pristine material and accretion inflows in it. Chen et al~\cite{Chen_2021} detected infalling streamers toward a high-mass YSO G352.63, with the temperature of the infalling gas $\sim 10$~K. It is very probable that the same inflow exist in S255IR and its part is projected onto the central part of SMA1.

\section{Optical depths of rotational lines.}
\label{sec:opt_depth}
The optical depth of a rotational transition can be estimated when individual hyperfine components of this transition are observed. We detected several hyperfine components of the $2-1$ transition of CN radical. It turned out that some of them are either blended with other lines or are weak, and five components of this transition are suitable for determining the optical depth of the CN lines. Then, it was discovered that the $F=1/2-1/2$ component at a frequency of 226663.685 MHz is anomalously strong. Perhaps it is blended with some unidentified line; the presence of hyperfine anomalies is also possible. We excluded this component and estimated the optical depth from the four remaining ones using the hfs method in the CLASS program. It turned out that the lines are optically thin, i.e., the optical depth of the most intense component does not exceed 0.1. However, it should be noted that such a procedure is incorrect in the case of hyperfine structure anomalies, and the obtained results should be used with caution.

The optical depth of molecular lines can also be estimated using substituted isotopologues. Suppose we observe the same transition of the primary and substituted isotopologues. In the case of equal and constant throughout the source excitation temperatures for both isotopologues the ratio of the line brightness temperatures is:
\begin{equation}
\frac{T^m_{br}}{T^s_{br}}=\frac{1-e^{-\tau}}{1-e^{-\tau/r}}
\end{equation}
where $T^m_{br}$ is the brightness temperature of the line of the main isotopologue, $T^s_{br}$ is the brightness temperature of the line of the substituted isotopologue, $\tau$ is the optical depth of the line of the main isotopologue, $r$ is the isotope ratio. Having determined the ratio $T^m_{br}/T^s_{br}$ from observations, and adopting a standard isotope ratio, one can find the values of the optical depths of the main and substituted isotopologues. Using this method we estimated optical depths of some molecular lines in SMA1 and SMA2.

\subsection{SMA1}
To determine the optical depths of methanol and formaldehyde lines, $^{13}$C isotopologues were used. The distance from the S255IR region to the Galactic center is 10.24 kpc; at this Galactocentric distance, the $^{12}$C/$^{13}$C ratio in the ISM is 82, according to formula (5) from \cite{Milam_2005}. In the case of methanol, the brightness temperatures of the $5_{-1}-4_{-1} \quad E$ lines of the main isotopologue and $^{13}$CH$_3$OH were compared. The optical depth of the $^{13}$CH$_3$OH line proved to be 0.29, and that of the $^{12}$CH$_3$OH line, $23.8\pm 1.5$. 

It should be noted that in the hot gas the abundance of most COMs with rotational partition functions comparable to or exceeding the partition function of methanol is one and a half to three orders of magnitude lower than the abundance of methanol (\cite{Arce_2008}, \cite{Kalenskii_2010}, \cite{Belloche_2013}, \cite{Kalenskii_2022}, \cite{Chen_2023} ). In addition, the $5_{-1}-4_{-1}E$ methanol line is one of the strongest among those we have detected; it is natural to assume that the characteristic values of the optical depths of other, weaker, methanol lines are lower than the optical depth of the $5_{-1}-4_{-1}E$ line. Since the abundance of the remaining COMs in the hot gas is 30 or more times lower than the abundance of methanol, we assume that the emission in the COMs is optically thin and hence the optical depths do not affect the rotation diagrams.

In the case of formaldehyde, the $3_{1,2}-2_{1,1}$ lines were used. The optical depth of the H$_2^{13}$CO line was equal to 0.15, and that of the H$_2$CO was $12.3\pm 1.0$.

Optical depth of the 5--4 CS line can be estimated using either the $^{13}$CS isotopologue or the C$^{34}$S isotopologue. We chose C$^{34}$S because the 5--4 $^{13}$CS line is blended with some unidentified line. The ratio $^{32}$S/$^{34}$S in the ISM is 22, according to~\cite{Wilson_1994}. Optical depth of the CS line proved to be $4.3\pm 0.5$, and the optical depth of the C$^{34}$S line proved to be 0.2.

In the case of SO, the $N,J=5,6-4,5$ transition of the main and  $^{34}$SO isotopologues were observed. Optical thickness of the line of the main isotopologue was found to be $2.70\pm 0.5$ and that of the $^{34}$SO isotopologue, 0.12. 

We recorded three lines of carbonyl sulfide (OCS) in SMA1--- 18--17, 19--18 and 20--19. Optical depths of the lines 18--17 and 20--19 were estimated using the corresponding lines of the OC$^{34}$S isotopologue. Optical depth of the 18--17 OCS line was $2.2\pm 1.2$, and that of the 20--19 OCS line, $1.4\pm 0.3$. Optical depths of the corresponding OC$^{34}$S lines were 0.1 and 0.06, respectively.
  
\subsection{SMA2}
In SMA2, we could estimate the optical depths of the 5--4 CS and $N,J=5,6-4,5$ SO lines. The optical depth of the CS line was determined using the $^{13}$CS and C$^{34}$S isotopologues. In the first case, it was found to be equal to $6.4\pm 0.5$, and in the second, $3.2\pm 0.2$. Unlike in SMA1, in SMA2 the parameters of the 5--4 lines of both the $^{13}$CS isotopologue and the C$^{34}$S isotopologue were determined reliably. The discrepancy between the obtained optical depth values may mean that either the [13C/12C] ratio, or the [34S/32S] ratio, or both of them differ from the accepted values. We used the average value of $4.8\pm 1.6$ as the optical depth of the 5-4 CS line.

Optical depth of the SO line was determined using the $^{34}$SO isotopologue; it was found to be $1.1\pm 0.2$.
  
\section{The origin of the ring-shaped structure}
\label{sec:discussion}

\begin{figure}
    \centering
\caption{Ring-shaped structure. The background represents the continuum emission at 1.3 mm. The white-red contours show the CN image in the 2-1 J=3/2-1/2 F=5/2-3/2 line at 226659.558~MHz; the contour levels are 1.34, 1.59, 1.85, 2.10, 2.36 and 2.61 Jy/beam. The green contours show the $5-4$ HCS$^+$ image at 213360.641 MHz; the contour levels are 0.62, 0.83 and 1.04 Jy/beam. The blue contours show the CH$_3$OH image in the $4_{2}-3_{1}E$ line at 218440.05~MHz; the contour levels are 4.31, 6.46, 8.61 and 10.76 Jy/beam.}
    \includegraphics[width=0.8\textwidth]{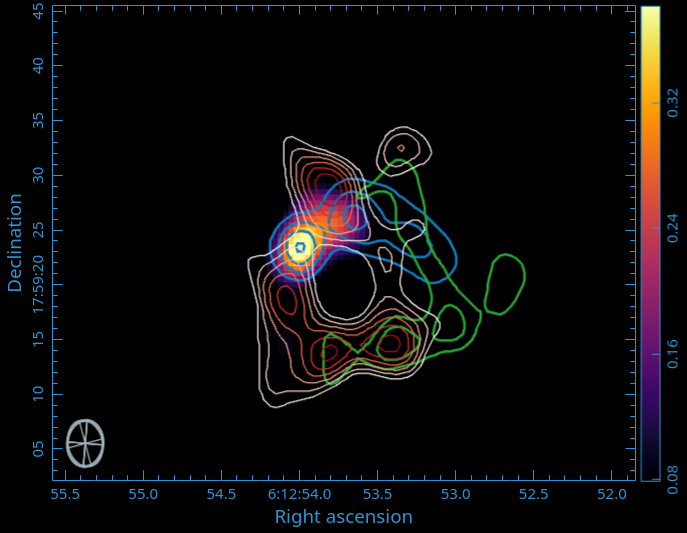}
    \label{fig:ring}
\end{figure}
One of the interesting features of the S255IR region detected as a result of the observations is the ring-shaped structure visible in the lines of CN and some other molecules (see~\ref{sec:results} and Fig.~\ref{fig:ring}) to the southeast of the SMA1--SMA3 region. It should be noted that no molecular emission was detected inside the ring, except for the weak emission of CO and its isotopologues $^{13}$CO and C$^{18}$O. A structure similar to CN image in Fig.~\ref{fig:ring} was partially observed previously in CCH and C$^{34}$S lines (see Figs. 4 and 6 in~\cite{Zinchenko_2020}). In addition, the contours of the $4_2-3_1$ methanol line in Fig.~\ref{fig:ring} roughly coincide with the CO and SiO images of the blue wing of the outflow from SMA2, presented in~\cite{Zinchenko_2020}.

It is natural to assume that the ring is formed by the walls of the cavity, previously discovered by the observations of the light echo from the flare in NIRS3 (\cite{Caratti_o_Garatti_2017}). According to \cite{Caratti_o_Garatti_2017}, this cavity is formed by the blue wing of the high-velocity outflow from NIRS3. However, it is possible that in addition to the NIRS3 outflow, the outflow from SMA2 (which at large distances may merge with the NIRS3 outflow) contributes to the formation of the cavity (\cite{Zinchenko_2020}). Probably there is one more outflow in this region, accelerated by NIRS1 and propagating perpendicular to the line of sight (\cite{Simpson_2009}, \cite{Zinchenko_2020}).

\begin{figure}
\caption{Rotation diagrams of methanol toward Met1 (left column) and Met2 (right column). Top: RDs of type I. Bottom: RDs of type II.}
\label{fig:rot_met_sma2}
\includegraphics[width=0.9\textwidth]{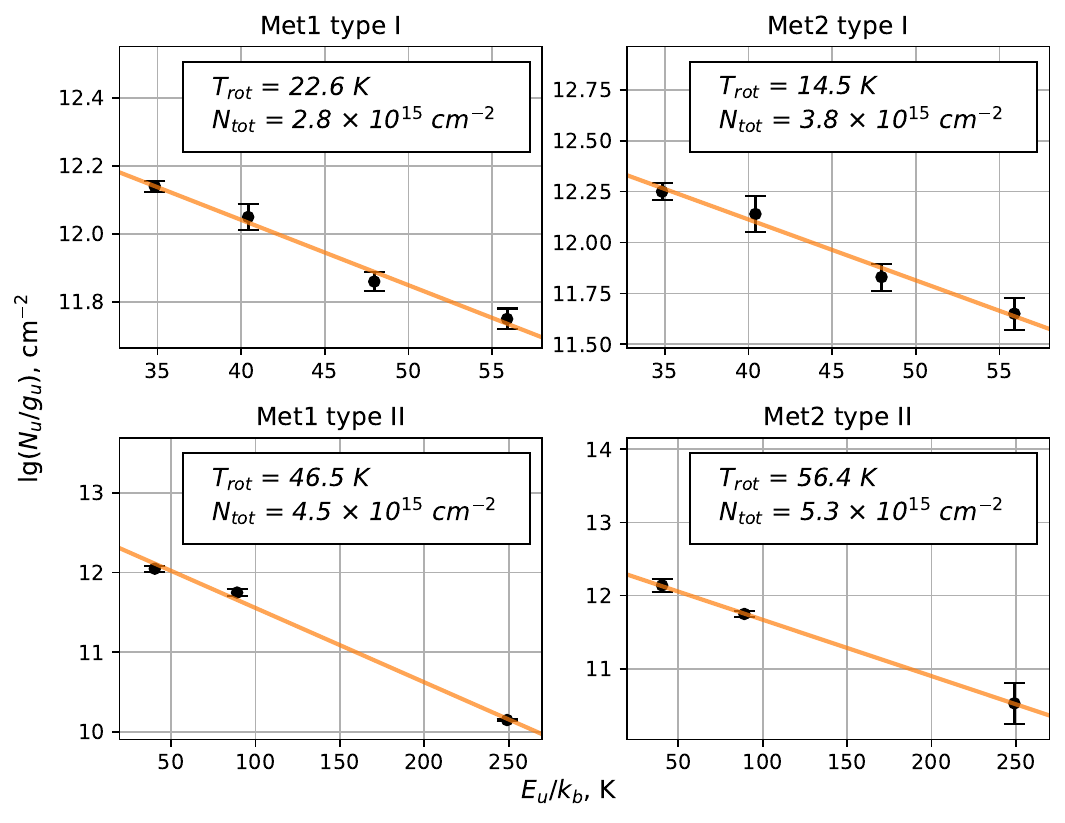}
\end{figure}

It is well known that many COMs are formed in the mantles of interstellar dust grains and enter the gas phase in large quantities during the evaporation or destruction of the mantles. Mantles evaporate when dust in the vicinity of protostars heats up to the temperature $\gtrsim 100$~K. The immediate vicinity of protostars with temperatures of the order of 100 K and higher, as well as with an increased content of complex molecules, are called hot cores. In hot cores, in particular, the content of oxygen-containing complex molecules (O-COMs) CH$_3$OH, HCOOH, CH$_3$CHO, CH$_3$OCHO, CH$_3$OCH$_3$ is greatly increased (e.g. Herbst \& van Dishoeck~\cite{Herbst_2009}). It is therefore not surprising that in S255IR these molecules are observed toward the hot core SMA1.

In addition, the abundance of such molecules can increase in hot gas behind the shock fronts of high-velocity outflows. In these sources, the mantles of dust grains are destroyed by collisions with fast H$_2$ molecules in hot gas. In S255IR, shock waves should arise at the points where high-velocity outflows interact with the cavity walls. It is well known that in some such objects the abundance of methanol is increased by several orders of magnitude compared to its abundance in quiescent gas (\cite{Bachiller_1995}, \cite{Bachiller_1997}), which leads to the appearance of strong emission in methanol lines. In addition, Arce et al.~\cite{Arce_2008} reported the detection of HCOOH, CH$_3$OH, CH$_3$OCHO, and CH$_3$CH$_2$OH toward the blue wing of the high-velocity outflow from the low-mass YSO L1157, as well as  measuring the abundance of these molecules relative to methanol. However, in S255IR,  of the four molecules reported by Arce et al., we only detected methanol in the direction of the high-velocity outflows associated with SMA1 and SMA2 (see Fig.~\ref{fig:multimap},\ref{fig:ch3oh}). Hence, the question arises, whether the non-detection of these molecules is because their abundance relative to methanol in S255IR is lower than in L1157, or because our sensitivity is not sufficient to detect their emission?

To answer this question, we first estimated the temperatures, the number densities of H$_2$ molecules, and the column densities of methanol toward two brightness peaks of methanol, designated Met1 and Met2~(see Fig.~\ref{fig:ch3oh}). Estimates were made using RDs constructed from specifically selected CH$_3$OH lines. According to Kalenskii and Kurtz~\cite{Kalenskii_2016}, an RD constructed using lines whose upper levels have the same rotational quantum number $J$, but different quantum numbers of the projection of the angular momentum of the molecule onto the symmetry axis $K$, allows one to determine the concentration of H$_2$ molecules rather than  source temperature. Kalenskii and Kurtz~\cite{Kalenskii_2016} denoted such RDs as type I RDs. The rotational temperature obtained using type I RD can be very different from the ambient temperature, but is uniquely related to the number density of H$_2$. If the rotation diagram is constructed using lines whose upper levels belong to the same $K$-ladder, i.e. have the same quantum number $K$, but different quantum numbers $J$, then the rotational temperature will be close to the gas temperature. Kalensky and Kurtz denoted such RDs as type II RDs. Figure~\ref{fig:rot_met_sma2} shows type I and II RDs toward Met1 and Met2. Type I RDs were constructed using the $5_K-4_K$ lines. The obtained rotational temperatures ($\gtrsim 15$~K) correspond to H$_2$ number densities $\gtrsim 10^7$~cm$^{-3}$; at such a high number density, the ratios of level populations are close to the values obtained as a result of collision thermalization, and the obtained methanol column densities ($N_{\rm CH_3OH}$) are close to the real ones, according to Table~1 from~\cite{Kalenskii_2016}. 

Type II RDs are constructed from the methanol-$E$ lines with the upper levels belonging to a ladder with quantum number $K=-1$ (the backbone ladder). The $T_{rot}$ and $N_{CH_3OH}$ values yielded by such RD, at high concentrations of H$_2$ molecules, are practically equal to the kinetic temperature of the source and the column density of methanol. Therefore, we assume that in Met1 the gas temperature is 47~K and the column density of methanol is $4.5\times 10^{15}$~cm$^{-2}$, and in Met2 these values are 56~K and $5.3\times 10^{15}$~cm$^{-2}$.

\begin{table}[htbp]
\centering
\caption{Column densities of O-COMs in Met1 and Met2. Columns 3 and 5 show column densities of undetected here molecules HCOOH, CH$_3$OCHO, and CH$_3$CH$_2$OH, estimated based on methanol column densities in these objects. Columns 4 and 6 show the minimum values of column densities of these molecules required for them to be detectable with our sensitivity. The values are calculated assuming LTE at 50~K}
\label{tab:coldens}
\begin{tabular}{|l|l|l|l|l|l|}
\hline
Molecule       & N$_{\rm mol}$/N$_{\rm CH_3OH}^1$ 
         &\multicolumn{2}{c|}{Met1}&\multicolumn{2}{c|}{Met2} \\   
\hline
               &           & N$_{mol}$, cm$^{-2}$ & N$_{mol}^{thr}$, cm$^{-2}$     & N$_{mol}$, cm$^{-2}$ & N$_{mol}^{thr}$, cm$^{-2}$\\
\hline
HCOOH          & $10^{-3}$ & $4.5\times 10^{12}$ & $5.2\times 10^{13}$ & $5.3\times 10^{12}$ & $5.0\times 10^{13}$\\
CH$_3$OCHO     & $10^{-2}$ & $4.5\times 10^{13}$ & $1.3\times 10^{14}$ & $5.3\times 10^{13}$ & $1.2\times 10^{14}$\\
CH$_3$CH$_2$OH & $10^{-3}$ & $4.5\times 10^{12}$ & $4.9\times 10^{14}$ & $5.3\times 10^{12}$ & $4.3\times 10^{14}$\\
\hline
\end{tabular}

\flushleft
$^1$--from Arce et al, 2008~\cite{Arce_2008}\\
\end{table}

Assuming that the abundances of HCOOH, CH$_3$OCHO, and CH$_3$CH$_2$OH relative to methanol are equal to those obtained by Arce et al.~\cite{Arce_2008}, we calculated column densities of these molecules in Met1 and Met2. The results are presented in Table~\ref{tab:coldens} (third and fifth columns). Assuming that the rotational temperatures of HCOOH, CH$_3$OCHO, and CH$_3$CH$_2$OH coincide with the rotational temperatures of methanol, we calculated, assuming LTE, the column densities of these molecules that would make them detectable with our sensitivity. We denote these values $N^{thr}_{mol}$; they are given in columns 4 and 6 of Table~\ref{tab:coldens}. Comparison of the values in the third column with those in the fourth column and in the fifth column with those in the sixth column shows that at our sensitivity we would not detect the emission of these complex oxygen-containing molecules toward the wings of the high-velocity outflows in S255IR. In subsequent work, we expect to decrease the detection thresholds for complex molecules using spectral line stacking and either detect O-COMs in Met1 and Met2 or show that the O-COM to methanol abundance ratios in S255IR are lower than in L1157.


\section{Summary and conclusion}
Using the SMA antenna array we performed interferometric observations of the star-forming region S255IR in the 210--250 GHz frequency range. As a result, emission of 53 molecules was detected, including COMs CH$_3$OCHO, CH$_3$OCH$_3$, CH$_3$CH$_2$OH, and numerous others. Maps of the region were built for a large number of molecular lines, and spectra of the SMA1 and SMA2 cores were obtained.

Most of the detected molecules can be roughly divided into two groups. Molecules of the first group emit exclusively toward the hot core SMA1, while molecules of the second group, in addition to SMA1, emit toward a ring-shaped structure west of SMA1. This structure appears to be associated with the walls of a cavity formed by high-velocity outflows from the YSOs in molecular cores SMA1, SMA2, and possibly SMA3. Brightness peaks of methanol, clearly visible toward the ring-shaped structure, may be associated with shock waves resulting from  the interaction between the outflows and the cavity walls.

We constructed rotation diagrams for 13 COMs detected in SMA1. Rotational temperatures for most of these molecules fall in the range of 100-200 K. Higher rotational temperatures were found for HCOOH ($T_{rot}$ = 359 K) and CH$_3$CN ($T_{rot}$ = 279 K). CH$_3$CH$_2$CN has a lower rotational temperature ($T_{rot}$ = 36 K). Methyl acetylene, which is considered a good ''thermometer'' for interstellar gas, has a negative rotational temperature, which likely indicates absorption of the K=0 and 1 emission in a cold shell.

Optical depths of the lines of several molecules in the SMA1 and SMA2 cores were estimated. In SMA1, the optical depth in the $5_{-1}-4_{-1}E$ methanol line proved to be $23.8\pm 1.5$. Since numerous studies show that the abundance of COMs such as CH$_3$CHO, CH$_3$OCHO, or C$_2$H$_5$OH in the hot gas is 30 or more times lower than the abundance of methanol, we assume that the emission of these COMs is optically thin and hence the optical depths do not affect the rotation diagrams. The optical depth in the $3_{1,2}-2_{1,2}$ formaldehyde line was found to be $12.3\pm 1.0$, in the $5-4$ CS line, $4.3\pm 0.5$, in the $N,J=5,6-4,5$ SO line, $2.70\pm 0.5$, and in the $18-17$ and $20-19$ OCS lines, $2.2\pm 1.2$ and $1.4\pm 0.3$, respectively.

In SMA2, the optical depth of the $5-4$ CS line was found to be $4.8\pm 1.6$, and the $N,J = 5,6-4,5$ SO line, $1.1\pm 0.2$.

Using methanol lines, gas temperature and density were estimated toward two objects seen as methanol brightness peaks Met1 and Met2. The temperature and density were found to be on the order of 50-60~K and $10^7-10^8$~cm$^{-3}$. Methanol column density in Met1 was found to be $5.4\times 10^{15}$ cm$^{-2}$, and in Met2 --- $4.0\times 10^{15}$ cm$^{-2}$. Column densities of other COMs in the ring-shaped structure will be determined in future studies with increased sensitivity using spectral line stacking.


\begin{acknowledgments}
The authors thanks prof. I. I. Zinchenko, S. V. Salii and P. A. Tanatova for useful discussions and to the anonymous referee for helpful comments. They are grateful for the support from the SMA operators during the observations.

 The Submillimeter Array is a joint project between the Smithsonian Astrophysical Observatory and the Academia Sinica Institute of Astronomy and Astrophysics and is funded by the Smithsonian Institution and the Academia Sinica.

We recognize that Maunakea is a culturally important site for the indigenous Hawaiian people; we are privileged to study the cosmos from its summit.
\end{acknowledgments}

\section*{Conflict of Interest}

The authors declare no conflict of interest.

\clearpage

\bibliographystyle{maik}
\bibliography{TemplateAR}

@article{Minier_2005,
   title={Star-forming protoclusters associated with methanol masers},
   volume={429},
   ISSN={1432-0746},
   url={http://dx.doi.org/10.1051/0004-6361:20041137},
   DOI={10.1051/0004-6361:20041137},
   number={3},
   journal={Astronomy \& Astrophysics},
   publisher={EDP Sciences},
   author={Minier, V. and Burton, M. G. and Hill, T. and Pestalozzi, M. R. and Purcell, C. R. and Garay, G. and Walsh, A. J. and Longmore, S.},
   year={2005},
   month=jan, pages={945–960} }

@ARTICLE{Snell_1986,
       author = {{Snell}, R.~L. and {Bally}, J.},
        title = "{Compact Radio Sources Associated with Molecular Outflows}",
      journal = {\apj},
         year = 1986,
        month = apr,
       volume = {303},
        pages = {683},
          doi = {10.1086/164117},
       adsurl = {https://ui.adsabs.harvard.edu/abs/1986ApJ...303..683S}
}

@article{Wang_2011,
   title={Different evolutionary stages in the massive star-forming region S255 complex},
   volume={527},
   ISSN={1432-0746},
   url={http://dx.doi.org/10.1051/0004-6361/201015543},
   DOI={10.1051/0004-6361/201015543},
   journal={Astronomy \& Astrophysics},
   publisher={EDP Sciences},
   author={Wang, Y. and Beuther, H. and Bik, A. and Vasyunina, T. and Jiang, Z. and Puga, E. and Linz, H. and Rodón, J. A. and Henning, Th. and Tamura, M.},
   year={2011},
   month=jan, pages={A32} }

@article{Zinchenko_2015,
   title={THE DISK-OUTFLOW SYSTEM IN THE S255IR AREA OF HIGH-MASS STAR FORMATION},
   volume={810},
   ISSN={1538-4357},
   url={http://dx.doi.org/10.1088/0004-637X/810/1/10},
   DOI={10.1088/0004-637x/810/1/10},
   number={1},
   journal={The Astrophysical Journal},
   publisher={American Astronomical Society},
   author={Zinchenko, I. and Liu, S.-Y. and Su, Y.-N. and Salii, S. V. and Sobolev, A. M. and Zemlyanukha, P. and Beuther, H. and Ojha, D. K. and Samal, M. R. and Wang, Y.},
   year={2015},
   month=aug, pages={10} }

@article{Boley_2013,
   title={The VLTI/MIDI survey of massive young stellar objects: Sounding the inner regions around intermediate- and high-mass young stars using mid-infrared interferometry⋆⋆⋆⋆⋆⋆},
   volume={558},
   ISSN={1432-0746},
   url={http://dx.doi.org/10.1051/0004-6361/201321539},
   DOI={10.1051/0004-6361/201321539},
   journal={Astronomy \& Astrophysics},
   publisher={EDP Sciences},
   author={Boley, Paul A. and Linz, Hendrik and van Boekel, Roy and Henning, Thomas and Feldt, Markus and Kaper, Lex and Leinert, Christoph and Müller, André and Pascucci, Ilaria and Robberto, Massimo and Stecklum, Bringfried and Waters, L. B. F. M. and Zinnecker, Hans},
   year={2013},
   month=sep, pages={A24} }

@article{Stecklum_2016,
  title={The methanol maser flare of S255IR and an outburst from the high-mass YSO S255IR-NIRS3 - more than a coincidence?},
 journal =  {The Astronomer's Telegram, No. 8732},
  author={Bringfried Stecklum; and Alessio Caratti o Garatti; and Maria Concepcion Cardenas; and Jochen Greiner; and Thomas Kruehler; and Sylvio Klose; and Jochen Eisloeffel},
  year={2016},
  url={https://www.astronomerstelegram.org/?read=8732}
}

@article{burns2016h2o,
  title={H2O masers in a jet-driven bow shock: episodic ejection from a massive young stellar object},
  author={Burns, RA and Handa, T and Nagayama, T and Sunada, K and Omodaka, T},
  journal={Monthly Notices of the Royal Astronomical Society},
  volume={460},
  number={1},
  pages={283--290},
  year={2016},
  publisher={Oxford University Press}
}

@article{wang2019elstag,
author = {Wang, Yuan and Beuther, H. and Bik, A. and Vasyunina, T. and Jiang, Z. and Puga, Erick and Linz, Haydenson and Rodon, J. and Henning, Th and Tamura, M.},
year = {2010},
month = {11},
pages = {},
title = {Different Evolutionary Stages in the Massive Star Forming Region S255
Complex},
volume = {527},
journal = {Astronomy and Astrophysics},
doi = {10.1051/0004-6361/201015543}
}

@article{Fujisawa2015AFO,
  title={A flare of methanol maser in S255},
  author={Kenta Fujisawa and Yoshinori Yonekura and Koichiro Sugiyama and Hikari Horiuchi and Takehiro Hayashi and Kazuya Hachisuka and Naoko Matsumoto and Kotaro Niinuma},
  year={2015},
journal = {The Astronomer's Telegram No. 8286},
  url={https://www.astronomerstelegram.org/?read=8286}
}

@article{Zinchenko_2020,
   title={Dense Cores, Filaments, and Outflows in the S255IR Region of High-mass Star Formation},
   volume={889},
   ISSN={1538-4357},
   url={http://dx.doi.org/10.3847/1538-4357/ab5c18},
   DOI={10.3847/1538-4357/ab5c18},
   number={1},
   journal={The Astrophysical Journal},
   publisher={American Astronomical Society},
   author={Zinchenko, Igor I. and Liu, Sheng-Yuan and Su, Yu-Nung and Wang, Kuo-Song and Wang, Yuan},
   year={2020},
   month=jan, pages={43} }

@article{Caratti_o_Garatti_2017,
   title={Disk-mediated accretion burst in a high-mass young stellar object},
   volume={13},
   ISSN={1745-2481},
   url={http://dx.doi.org/10.1038/nphys3942},
   DOI={10.1038/nphys3942},
   number={3},
   journal={Nature Physics},
   publisher={Springer Science and Business Media LLC},
   author={Caratti o Garatti, A. and Stecklum, B. and Garcia Lopez, R. and Eislöffel, J. and Ray, T. P. and Sanna, A. and Cesaroni, R. and Walmsley, C. M. and Oudmaijer, R. D. and de Wit, W. J. and Moscadelli, L. and Greiner, J. and Krabbe, A. and Fischer, C. and Klein, R. and Ibañez, J. M.},
   year={2017},
   month=mar, pages={276–279} }

@article{Uchiyama_2019,
   title={Near-infrared monitoring of the accretion outburst in the massive young stellar object S255-NIRS3},
   volume={72},
   ISSN={2053-051X},
   url={http://dx.doi.org/10.1093/pasj/psz122},
   DOI={10.1093/pasj/psz122},
   number={1},
   journal={Publications of the Astronomical Society of Japan},
   publisher={Oxford University Press (OUP)},
   author={Uchiyama, Mizuho and Yamashita, Takuya and Sugiyama, Koichiro and Nakaoka, Tatsuya and Kawabata, Miho and Itoh, Ryosuke and Yamanaka, Masayuki and Akitaya, Hiroshi and Kawabata, Koji and Yonekura, Yoshinori and Saito, Yu and Motogi, Kazuhito and Fujisawa, Kenta},
   year={2019},
   month=nov }

@article{Liu_2018,
   title={A Submillimeter Burst of S255IR SMA1: The Rise and Fall of Its Luminosity},
   volume={863},
   ISSN={2041-8213},
   url={http://dx.doi.org/10.3847/2041-8213/aad63a},
   DOI={10.3847/2041-8213/aad63a},
   number={1},
   journal={The Astrophysical Journal Letters},
   publisher={American Astronomical Society},
   author={Liu, Sheng-Yuan and Su, Yu-Nung and Zinchenko, Igor and Wang, Kuo-Song and Wang, Yuan},
   year={2018},
   month=aug, pages={L12} }

@inproceedings{Zinchenko_2024,
   title={Study of the high-mass star-forming region S255IR at various scales},
   url={http://dx.doi.org/10.26119/VAK2024.096},
   DOI={10.26119/vak2024.096},
   booktitle={Modern astronomy: from the Early Universe to exoplanets and black holes},
   publisher={Special Astrophysical Observatory of the Russian Academy of Sciences},
   author={Zinchenko, Igor and Liu, S. and Ojha, D. and Su, Y. and Zemlyanukha, P.},
   year={2024},
   month=dec, pages={604–610} }

@article{Zinchenko_2012,
   title={A MULTI-WAVELENGTH HIGH-RESOLUTION STUDY OF THE S255 STAR-FORMING REGION: GENERAL STRUCTURE AND KINEMATICS},
   volume={755},
   ISSN={1538-4357},
   url={http://dx.doi.org/10.1088/0004-637X/755/2/177},
   DOI={10.1088/0004-637x/755/2/177},
   number={2},
   journal={The Astrophysical Journal},
   publisher={American Astronomical Society},
   author={Zinchenko, I. and Liu, S.-Y. and Su, Y.-N. and Kurtz, S. and Ojha, D. K. and Samal, M. R. and Ghosh, S. K.},
   year={2012},
   month=aug, pages={177} }

@article{Fedriani_2023,
   title={The sharpest view on the high-mass star-forming region S255IR: Near infrared adaptive optics imaging of the outbursting source NIRS3},
   volume={676},
   ISSN={1432-0746},
   url={http://dx.doi.org/10.1051/0004-6361/202346736},
   DOI={10.1051/0004-6361/202346736},
   journal={Astronomy \& Astrophysics},
   publisher={EDP Sciences},
   author={Fedriani, R. and Caratti o Garatti, A. and Cesaroni, R. and Tan, J. C. and Stecklum, B. and Moscadelli, L. and Koutoulaki, M. and Cosentino, G. and Whittle, M.},
   year={2023},
   month=aug, pages={A107} }

@article{Simpson_2009,
   title={HUBBLE SPACE TELESCOPENICMOS POLARIZATION OBSERVATIONS OF THREE EDGE-ON MASSIVE YOUNG STELLAR OBJECTS},
   volume={700},
   ISSN={1538-4357},
   url={http://dx.doi.org/10.1088/0004-637X/700/2/1488},
   DOI={10.1088/0004-637x/700/2/1488},
   number={2},
   journal={The Astrophysical Journal},
   publisher={American Astronomical Society},
   author={Simpson, Janet P. and Burton, Michael G. and Colgan, Sean W. J. and Cotera, Angela S. and Erickson, Edwin F. and Hines, Dean C. and Whitney, Barbara A.},
   year={2009},
   month=jul, pages={1488–1501} }

@article{Liu_2020,
   title={ALMA View of the Infalling Envelope around a Massive Protostar in S255IR SMA1},
   volume={904},
   ISSN={1538-4357},
   url={http://dx.doi.org/10.3847/1538-4357/abc0ec},
   DOI={10.3847/1538-4357/abc0ec},
   number={2},
   journal={The Astrophysical Journal},
   publisher={American Astronomical Society},
   author={Liu, Sheng-Yuan and Su, Yu-Nung and Zinchenko, Igor and Wang, Kuo-Song and Meyer, Dominique M.-A. and Wang, Yuan and Hsieh, I-Ta},
   year={2020},
   month=dec, pages={181} }

@ARTICLE{Milam_2005,
       author = {{Milam}, S.~N. and {Savage}, C. and {Brewster}, M.~A. and {Ziurys}, L.~M. and {Wyckoff}, S.},
        title = "{The $^{12}$C/$^{13}$C Isotope Gradient Derived from Millimeter Transitions of CN: The Case for Galactic Chemical Evolution}",
      journal = {\apj},
         year = 2005,
        month = dec,
       volume = {634},
       number = {2},
        pages = {1126-1132},
          doi = {10.1086/497123},
       adsurl = {https://ui.adsabs.harvard.edu/abs/2005ApJ...634.1126M}
}

@ARTICLE{Keating2025,
       author = {{Keating}, Garrett and {Hazelton}, Bryna and {Kolopanis}, Matthew and {Murray}, Steven and {Beardsley}, Adam and {Jacobs}, Daniel and {Kern}, Nicholas and {Lanman}, Adam and {La Plante}, Paul and {Pober}, Jonathan and {Star}, Pyxie},
        title = "{pyuvdata v3: an interface for astronomical interferometric data sets in Python}",
      journal = {The Journal of Open Source Software},
         year = 2025,
        month = may,
       volume = {10},
       number = {109},
          eid = {7482},
        pages = {7482},
          doi = {10.21105/joss.07482},
       adsurl = {https://ui.adsabs.harvard.edu/abs/2025JOSS...10.7482K}
}

@ARTICLE{Hazelton2017,
       author = {{Hazelton}, Bryna J. and {Jacobs}, Daniel C. and {Pober}, Jonathan C. and {Beardsley}, Adam P.},
        title = "{pyuvdata: an interface for astronomical interferometeric datasets in python}",
      journal = {The Journal of Open Source Software},
         year = 2017,
        month = feb,
       volume = {2},
       number = {10},
          eid = {140},
        pages = {140},
          doi = {10.21105/joss.00140},
       adsurl = {https://ui.adsabs.harvard.edu/abs/2017JOSS....2..140H}
}

@article{Bachiller_1995,
   title={Methanol enhancemant in young bipolar outflows},
   volume={295},
   ISSN={},
   url={https://articles.adsabs.harvard.edu/pdf/1995A%26A...295L..51B},
   DOI={},
   number={},
   journal={Astronomy and Astrophysics},
   publisher={},
   author={Bachiller, R. and Liechti, S. and Walmsley, C. M. and Colomer, F.},
   year={1995},
   month=mar, pages={L51–L54} }

@article{Bachiller_1997,
   title={Shock Chemistry in the Young Bipolar Outflow L1157},
   volume={487},
   ISSN={},
   url={https://iopscience.iop.org/article/10.1086/310877},
   DOI={10.1086/310877},
   number={1},
   journal={Astronomy and Astrophysics},
   publisher={},
   author={Bachiller, R. and Pérez Gutiérrez, M. },
   year={1997},
   month=sep, pages={L93–L96} }

@article{Arce_2008,
   title={Complex Molecules in the L1157 Molecular Outflow},
   volume={681},
   ISSN={1538-4357},
   url={http://dx.doi.org/10.1086/590110},
   DOI={10.1086/590110},
   number={1},
   journal={The Astrophysical Journal},
   publisher={American Astronomical Society},
   author={Arce, Héctor G. and Santiago-García, Joaquín and Jørgensen, Jes K. and Tafalla, Mario and Bachiller, Rafael},
   year={2008},
   month=jun, pages={L21–L24} }

@article{Kalenskii_2010,
   title={Spectral Survey of the Star-Forming Region W51 e1/e2 at 3mm},
   volume={54},
   ISSN={},
   url={},
   DOI={},
   number={12},
   journal={Astronomy Reports},
   publisher={},
   author={Kalenskii, S. V. and Johansson, L. E. B.},
   year={2010},
   month=dec, pages={1084-1104} }

@article{Kalenskii_2022,
   title={Spectral-line Survey of the Region of Massive Star Formation W51e1/e2 in the 4 mm Wavelength Range},
   volume={932},
   ISSN={1538-4357},
   url={},
   DOI={},
   number={},
   journal={The Astrophysical Journal},
   publisher={American Astronomical Society},
   author={Kalenskii, S. V. and Kaiser, R. I. and Bergman, P. and Olofsson, A. O. H. and Degtyarev, K. D. and Golysheva, P.},
   year={2022},
   month=jun, pages={} }

@article{Kalenskii_2016,
   title={Analytical Methods for Measuring the Parameters
of Interstellar Gas Using Methanol Observations},
   volume={60},
   ISSN={},
   url={},
   DOI={10.1134/S1063772916080047},
   number={8},
   journal={Astronomy Reports},
   publisher={Pleiades Publishing, Ltd.},
   author={Kalenskii, S. V. and Kurtz, S.},
   year={2016},
   month=aug, pages={702-717} }

@article{Wilson_1994,
   title={Abundances in the Interstellar Medium},
   volume={32},
   ISSN={},
   url={},
   DOI={https://doi.org/10.1146/annurev.aa.32.090194.001203},
   number={},
   journal={Annual Review of Astronomy and Astrophysics},
   publisher={},
   author={Wilson, T. L. and Rood, R. T.},
   year={1994},
   month=sep, pages={191-226} }

@article{Herbst_2009,
   title={Complex Organic Interstellar Molecules},
   volume={47},
   ISSN={},
   url={},
   DOI={https://doi.org/10.1146/annurev-astro-082708-101654},
   number={1},
   journal={Annual Review of Astronomy and Astrophysics},
   publisher={},
   author={Herbst, E. and van Dishoeck, E. F.},
   year={2009},
   month=sep, pages={427-480} }

@article{Belloche_2013,
   title={ Complex organic molecules in the interstellar medium: IRAM 30 m line survey of Sagittarius B2(N) and (M)},
   volume={559},
   ISSN={},
   url={},
   DOI={10.1051/0004-6361/201321096 },
   number={},
   journal={Astronomy \& Astrophysics},
   publisher={},
   author={Belloche, A. and Muller, H. S. P. and Menten, K. M. and Schilke, P. and Comito, C.},
   year={2013},
   month=nov, pages={47} }

@article{Chen_2023,
   title={CoCCoA: Complex Chemistry in hot Cores with ALMA},
   volume={678},
   ISSN={},
   url={},
   DOI={10.1051/0004-6361/202346491 },
   number={},
   journal={Astronomy \& Astrophysics},
   publisher={},
   author={Chen, Y. and van Gelder, M. L. and Nazari, P. and Brogan, C. L. and van Dishoeck, E. F. and Linnartz, H. and Jørgensen, J. K. and Hunter, T. R. and Wilkins, O. H. and Blake, G. A. and Caselli, P. and Chuang, K.-J. and Codella, C. and Cooke, I. and Drozdovskaya, M. N. and Garrod, R. T. and Ioppolo, S. and Jin, M. and Kulterer, B. M. and Ligterink, N. F. W. },
   year={2023},
   month=oct, pages={A137} }

@software{angus_comrie_2018_3377984,
author       = {Angus Comrie and
               Kuo-Song Wang and
               Yu-Hsuan Hwang and
               Anthony Moraghan and
               Pamela Harris and
               Adrianna Pińska and
               Carli Raul-Omar and
               Kuan-Chou Hou and
               Cheng-Chin Chiang and
               Tien-Hao Chang and
               Shou-Chieh Hsu and
               Qi Pang and
               Rob Simmonds and
               Po-Sheng Huang and
               Ming-Yi Lin and
               Hengtai Jan},
title        = {{CARTA: The Cube Analysis and Rendering Tool for
                Astronomy}},
month        = dec,
year         = 2018,
publisher    = {Zenodo},
doi          = {10.5281/zenodo.3377984},
url          = {https://doi.org/10.5281/zenodo.3377984}
}

@article{Chen_2021,
       author = {{Chen}, Xi and {Ren}, Zhi-Yuan and {Li}, Da-Lei and {Liu}, Tie and {Wang}, Ke and {Shen}, Zhi-Qiang and {Ellingsen}, Simon P. and {Sobolev}, Andrej M. and {Mei}, Ying and {Li}, Jing-Jing and {Wu}, Yue-Fang and {Kim}, Kee-Tae},
        title = "{Chemically Fresh Gas Inflows Detected in a Nearby High-mass Star-forming Region}",
      journal = {The Astrophysical Journal Letters},
         year = 2021,
        month = dec,
       volume = {923},
       number = {1},
          eid = {L20},
        pages = {L20},
          doi = {10.3847/2041-8213/ac3ec8},
       adsurl = {https://ui.adsabs.harvard.edu/abs/2021ApJ...923L..20C}
}

@article{Ojha_2011,
   title={Star formation activity in the Galactic Hii complex S255–S257},
   volume={738},
   ISSN={},
   url={},
   DOI={10.1088/0004-637X/738/2/156 },
   number={},
   journal={Astrophysical Journal},
   publisher={},
   author={Ojha, D. K. and Samal, M. R. and Pandey, A. K. and Bhatt, B. C. and Ghosh, S. K. and Sharma, S. and Tamura, M. and Mohan, V. and Zinchenko, I.},
   year={2011},
   month=sep, pages={156} }

@article{Goddi_2007,
   title={Associations of H$_2$O and CH$_3$OH masers at milli-arcsec angular resolution in two high-mass YSOs},
   volume={461},
   ISSN={},
   url={},
   DOI={10.1051/0004-6361:20066136 },
   number={},
   journal={Astronomy \& Astrophysics},
   publisher={},
   author={Goddi, C. and Moscadelli, L. and Sanna, A. and Cesaroni, R. and Minier, V.},
   year={2007},
   month={}, pages={1027-1035} }

@article{Minier_2000,
   title={VLBI observations of 6.7 and 12.2 GHz methanol masers toward high mass star-forming regions. I. Observational results: protostellar disks or outflows?},
   volume={362},
   ISSN={},
   url={},
   DOI={},
   number={},
   journal={Astronomy \& Astrophysics},
   publisher={EDP Sciences},
   author={Minier, V. and Booth, R. S. and Conway, J. E.},
   year={2000},
   month={}, pages={1093–1108} }

\clearpage
\section*{APPENDIX}
\begin{longtable}{|l|r|l|r|l|}
\caption{Spectral lines for which maps were constructed and which were used to build rotation diagrams. Comments: map — a map was constructed for this line; rd — the line was used for the rotation diagram.}
\label{tab:lines}\\
\hline
Molecule&Frequency&Transition& Low level         &Comments \\
         & (MHz)  &         & energy (cm$^{-1}$)&\\
\hline
\endfirsthead
\hline
Molecule&Frequency&Transition& Low level         &Comments \\
         & (MHz)  &         & energy (cm$^{-1}$)&\\
\hline
\endhead

\hline
\endfoot

\hline
\endlastfoot

$^{13}$CO       & 220398.684 & $2-1 $                   & 3.6759 &map \\
C$^{17}$O       & 224714.389 & $2-1$                    & 3.7479 &map \\
C$^{18}$O       & 219560.358 & $2-1$                    & 3.6619 &map \\
C$^{34}$S       & 241016.089 & $5-4$                    & 16.0796&map \\
CH$_2$CO        & 220177.416 & $11_{1,11}-10_{1,10}$ & 45.8108 &rd \\
CH$_2$CO        & 222228.587 & $11_{2,10}-10_{2,9}$ & 73.4814 &rd \\
CH$_2$CO        & 222314.457 & $11_{2,9}-10_{2,8}$ & 73.4878 &rd \\
CH$_2$CO        & 224327.250 & $11_{1,10}-10_{1,9}$ & 46.4891&map, rd \\
CH$_2$CO        & 240185.612 & $12_{1,12}-11_{1,11}$ & 53.1551 &rd \\
CH$_2$CO        & 242309.307 & $12_{4,9}-11_{4,8}$ & 191.2698 &rd \\
+CH$_2$CO        & 242309.308 & $12_{4,8}-11_{4,7}$ & 191.2698 &rd \\
CH$_2$CO        & 242375.380 & $12_{0,12}-11_{0,11}$ & 44.4798 &rd \\
CH$_2$CO        & 242398.458 & $12_{3,10}-11_{3,9}$ & 126.6753 &rd \\
+CH$_2$CO        & 242398.956 & $12_{3,9}-11_{3,8}$ & 126.6753 &rd \\
CH$_2$CO        & 242424.606 & $12_{2,11}-11_{2,10}$ & 80.8941 &rd \\
CH$_2$CO        & 244712.079 & $12_{1,11}-11_{1,10}$ & 53.9858 &rd \\
CH$_3$CCH       & 222128.812 & $13_{3}-12_{3}$ & 89.5037 &rd \\
CH$_3$CCH       & 222150.008 & $13_{2}-12_{2}$ & 64.4833 &rd \\
CH$_3$CCH       & 222162.729 & $13_{1}-12_{1}$ & 49.4710 &rd \\
CH$_3$CCH       & 222166.971 & $13_{0}-12_0$   & 44.4669 &map, rd \\
CH$_3$CCH       & 239211.212 & $14_{3}-13_{3}$ & 96.9131 &rd \\
CH$_3$CCH       & 239234.032 & $14_{2}-13_{2}$ & 71.8934 &rd \\
CH$_3$CCH       & 239247.727 & $14_{1}-13_{1}$ & 56.8816 &rd \\
CH$_3$CCH       & 239252.294 & $14_0-13_0$     & 51.8776&map, rd \\
CH$_3$CH$_2$CN  & 215039.727 & $24_{9,* }-23_{9,* }$ & 145.0122 &rd \\
+CH$_3$CH$_2$CN  & 215041.902 & $24_{10,* }-23_{10,* }$ & 159.6548 &rd \\
CH$_3$CH$_2$CN  & 215109.050 & $24_{7,* }-23_{7,* }$ & 120.3350 &rd \\
CH$_3$CH$_2$CN  & 215119.210 & $25_{0,25}-24_{0,24}$ & 87.2402 &rd \\
CH$_3$CH$_2$CN  & 215965.588 & $25_{1,25}-24_{0,24}$ & 87.2402 &rd \\
CH$_3$CH$_2$CN  & 215941.073 & $6_{4,3}-5_{3,2}$ &  11.4397 & rd \\
CH$_3$CH$_2$CN  & 222918.177 & $25_{1,24}-24_{1,23}$ & 91.4821 &rd \\
CH$_3$CH$_2$CN  & 223553.585 & $26_{0,26}-25_{0,25}$ & 94.4158 &rd \\
CH$_3$CH$_2$CN  & 225236.118 & $25_{4,21}-24_{4,20}$     &102.1946&map, rd \\
CH$_3$CH$_2$CN  & 225317.146 & $23_{2,22}-22_{1,21}$ & 77.4544 &rd \\
CH$_3$CH$_2$CN  & 227780.978 & $25_{3,22}-24_{3,21}$ & 97.2481 &rd \\
CH$_3$CH$_2$CN  & 231854.217 & $27_{1,27}-26_{1,26}$ & 101.8954 &rd \\
CH$_3$CH$_2$CN  & 235051.950 & $13_{3,10}-12_{2,11}$ &  26.3395 & rd \\
CH$_3$CH$_2$CN  & 241922.546 & $27_{10,* }-26_{10,* }$ & 182.0705 &rd \\
CH$_3$CH$_2$CN  & 245023.654 & $14_{3,11}-13_{2,12}$ & 30.2053 &rd \\
CH$_3$CH$_2$CN  & 248617.257 & $16_{3,14}-15_{2,13}$ &  39.3323 & rd \\
g-CH$_3$CH$_2$OH & 225109.881 & $13_{6,*}-12_{6,*} \nu_t=1-1$ & 118.6281 &rd \\
g-CH$_3$CH$_2$OH & 225170.614 & $13_{7,*}-12_{7,*} \nu_t=0-0$ & 126.1147 &rd \\
g-CH$_3$CH$_2$OH & 225248.812 & $13_{6,*}-12_{6,*} \nu_t=0-0$ & 115.0929 &rd \\
g-CH$_3$CH$_2$OH & 225399.733 & $13_{5,9}-12_{5,8} \nu_t=0-0$ & 105.7788 &rd \\
g-CH$_3$CH$_2$OH & 225404.089 & $13_{5,8}-12_{5,7} \nu_t=0-0$ & 105.7789 &rd \\
g-CH$_3$CH$_2$OH & 227760.743 & $3_{2,2}-2_{1,2} \nu_t=1-0$ & 42.0627 &rd \\
g-CH$_3$CH$_2$OH & 228029.050 & $13_{3,10}-12_{3,9} \nu_t=0-0$ & 92.4075 &rd \\
g-CH$_3$CH$_2$OH & 230793.506 & $6_{5,1}-5_{4,1} \nu_t=0-1$ & 65.1423 &rd \\
+g-CH$_3$CH$_2$OH & 230793.506 & $6_{5,2}-5_{4,2} \nu_t=0-1$ & 65.1423 &rd \\
+ AlF  & 230793.905  & $7-6$ & 23.098 & \\
g-CH$_3$CH$_2$OH & 231668.733 & $14_{1,14}-13_{1,13} \nu_t=0-0$ & 90.9002 &rd \\
g-CH$_3$CH$_2$OH & 232491.366 & $14_{0,14}-13_{0,13} \nu_t=0-0$ & 90.7911 &rd \\
g-CH$_3$CH$_2$OH & 232596.554 & $14_{1,14}-13_{0,13} \nu_t=1-1$ & 94.0893 &rd \\
g-CH$_3$CH$_2$OH & 238568.317 & $6_{1,5}-5_{0,5} \nu_t=1-0 $ & 48.1500 &rd \\
g-CH$_3$CH$_2$OH & 239478.079 & $14_{2,13}-13_{2,12} \nu_t=0-0 $ & 95.1822 &rd \\
g-CH$_3$CH$_2$OH & 239551.366 & $14_{2,13}-13_{2,12} \nu_t=1-1 $ & 98.4279 &rd \\
g-CH$_3$CH$_2$OH & 242349.842 & $14_{7,*}-13_{7,*} \nu_t=1-1 $ & 137.2373 &rd \\
g-CH$_3$CH$_2$OH & 242685.010 & $14_{5,10}-13_{5,9} \nu_t=1-1 $ & 116.7465 &rd \\
g-CH$_3$CH$_2$OH & 242770.099 & $14_{3,12}-13_{3,11} \nu_t=1-1 $ & 103.0917 &rd \\
g-CH$_3$CH$_2$OH & 243120.317 & $14_{4,11}-13_{4,10}$   & 105.7049 &map, rd \\
g-CH$_3$CH$_2$OH & 244633.950 & $14_{1,13}-13_{1,12} \nu_t=1-1 $ & 97.3154 &rd \\
g-CH$_3$CH$_2$OH & 245327.139 & $14_{3,11}-13_{3,10} \nu_t=1-1 $ & 103.2617 &rd \\
g-CH$_3$CH$_2$OH & 246414.762 & $14_{3,11}-13_{3,10} \nu_t=0-0$ & 100.0137 &rd \\
g-CH$_3$CH$_2$OH & 248178.548 & $15_{1,15}-14_{1,14} \nu_t=1-1 $ & 101.8479 &rd \\
g-CH$_3$CH$_2$OH & 248463.614 & $14_{2,12}-13_{2,11} \nu_t=0-0 $ & 96.3457 &rd \\
g-CH$_3$CH$_2$OH & 248577.119 & $15_{0,15}-14_{0,14} \nu_t=0-0 $ & 98.5462 &rd \\
t-CH$_3$CH$_2$OH & 218554.382 & $21_{5,16}-21_{4,17}$ & 149.7973 & rd \\
t-CH$_3$CH$_2$OH & 218654.008 & $7_{2,5}-6_{1,6}$ &  12.6749 & rd \\
t-CH$_3$CH$_2$OH & 225229.253 & $17_{2,15}-16_{3,14}$ &  87.3973 & rd \\
t-CH$_3$CH$_2$OH & 229491.13  & $17_{5,12}-17_{4,13}$ & 103.6309 & rd \\
t-CH$_3$CH$_2$OH & 230991.377 & $14_{0,14}-13_{1,13}$ &  51.7390 & rd \\
t-CH$_3$CH$_2$OH & 231737.593 & $19_{5,15}-19_{4,16}$ & 125.2481 & rd \\
t-CH$_3$CH$_2$OH & 231790.000 & $22_{5,18}-22_{4,19}$ & 162.2344 & rd \\
t-CH$_3$CH$_2$OH & 232034.630 & $18_{5,14}-18_{4,15}$ & 114.0935 & rd \\
t-CH$_3$CH$_2$OH & 232075.864 & $15_{5,10}-15_{4,11}$ &  84.2068 & rd \\
t-CH$_3$CH$_2$OH & 232318.469 & $23_{5,19}-23_{4,20}$ & 175.7316 & rd \\
t-CH$_3$CH$_2$OH & 232928.552 & $14_{5,9}-14_{4,10}$ &  75.3953 & rd \\
t-CH$_3$CH$_2$OH & 233571.082 & $13_{5,8}-13_{4,9}$ &  67.1801 & rd \\
t-CH$_3$CH$_2$OH & 233951.264 & $13_{5,9}-13_{4,10}$ &  67.1670 & rd \\
t-CH$_3$CH$_2$OH & 234051.178 & $12_{5,7}-12_{4,8}$ &  59.5587 & rd \\
t-CH$_3$CH$_2$OH & 234255.161 & $12_{5,8}-12_{4,9}$     &59.5518 &map, rd \\
t-CH$_3$CH$_2$OH & 234406.456 & $11_{5,6}-11_{4,7}$ &  52.5292 & rd \\
t-CH$_3$CH$_2$OH & 234666.157 & $10_{5,5}-10_{4,6}$ &  46.0900 & rd \\
t-CH$_3$CH$_2$OH & 234758.820 & $6_{3,4}-5_{2,3}$       &12.2822 &map, rd \\
t-CH$_3$CH$_2$OH & 234852.866 & $9_{5,4}-9_{4,5}$ &  40.2396 & rd \\
t-CH$_3$CH$_2$OH & 234873.877 & $9_{5,5}-9_{4,6}$ &  40.2389 & rd \\
t-CH$_3$CH$_2$OH & 234984.050 & $8_{5,3}-8_{4,4}$ &  34.9769 & rd \\
t-CH$_3$CH$_2$OH & 234992.183 & $8_{5,4}-8_{4,5}$ &  34.9766 & rd \\
t-CH$_3$CH$_2$OH & 235073.313 & $7_{5,2}-7_{4,3}$ &  30.3011 & rd \\
t-CH$_3$CH$_2$OH & 235131.372 & $6_{5,1}-6_{4,2}$ &  26.2113 & rd \\
t-CH$_3$CH$_2$OH & 236146.424 & $15_{1,14}-14_{2,13}$ &  64.3344 & rd \\
t-CH$_3$CH$_2$OH & 238667.995 & $27_{5,23}-27_{4,24}$ & 235.4995 & rd \\
t-CH$_3$CH$_2$OH & 243556.853 & $8_{2,6}-7_{1,7}$ &  16.5989 & rd \\
CH$_3$CHO     & 216581.924 & $11_{1,10}-10_{1,9}E$      &37.8632 &rd\\
CH$_3$CHO     & 223650.097 & $12_{-1,12}-11_{-1,11}E$   &42.7701 &rd\\
CH$_3$CHO     & 223660.610 & $12_{1,12}-11_{1,11}A^{++}$&42.7701 &rd\\
CH$_3$CHO     & 224656.971 & $12_{3,9}-12_{2,10}A^{-+}$ &56.8839 &rd\\
CH$_3$CHO     & 226487.236 & $13_{0,13}-12_{-1,12}E$    &50.2302 &rd\\
CH$_3$CHO     & 226551.622 &$12_{0,12}-11_{0,11}E$      &42.0628&map,rd\\
CH$_3$CHO     & 226592.725 &$12_{0,12}-11_{0,11}A^{++}$ &42.0013&map,rd\\
CH$_3$CHO     & 227258.988 & $12_{0,12}-11_{0,11}E~\nu_t=1$&184.0719 &rd\\
CH$_3$CHO     & 229432.106 & $11_{-1,11}-10_{0,10}E$    &35.1170 &rd\\
CH$_3$CHO     & 229775.029 & $11_{1,11}-10_{0,10}A^{++}$&35.0542 &rd\\
CH$_3$CHO     & 230301.924 & $12_{2,11}-11_{2,10}A^{--}$&48.6454 &rd\\
CH$_3$CHO     & 230315.788 & $12_{-2,11}-11_{-2,10}E$   &48.6543 &rd\\
CH$_3$CHO     & 230395.155 & $12_{2,11}-11_{2,10}A^{--}~\nu_t=1$&191.378 &rd\\
CH$_3$CHO     & 231363.289 & $12_{-5,7}-11_{-5,6}E$     &81.6239 &rd\\
CH$_3$CHO     & 231369.834 & $12_{5,8}-11_{5,7}E$       &81.6023 &rd\\
CH$_3$CHO     & 231456.738 & $12_{4,9}-11_{4,8}A^{--}$  &67.5887 &rd\\
CH$_3$CHO     & 231467.499 & $12_{4,8}-11_{4,7}A^{++}$  &67.5891 &rd\\
CH$_3$CHO     & 231506.297 & $12_{-4,9}-11_{-4,8}E$     &67.5167 &rd\\ 
CH$_3$CHO     & 231595.269 & $12_{3,10}-11_{3,9}A^{++}$ &56.6212 &rd\\ 
CH$_3$CHO     & 231748.722 & $12_{-3,10}-11_{-3,9}E$    &56.5679 &rd\\ 
CH$_3$CHO     & 231847.575 & $12_{3,9}-11_{3,8}E$       &56.6340 &rd\\ 
CH$_3$CHO     & 234795.450 & $12_{2,10}-11_{2,9}E$      &49.0668 &rd\\
CH$_3$CHO     & 234825.872 &$12_{2,10}-11_{2,9}A^{++}$  &49.0510&map,rd\\
CH$_3$CHO     & 234842.78  & $6_{3,3}-6_{2,4}A$         &19.8344 &rd\\
CH$_3$CHO     & 235217.832 &$12_{1,11}-11_{1,10}A^{--}~\nu_t=1$&188.3389 &rd\\
CH$_3$CHO     & 235996.212 &$12_{1,11}-11_{1,10}E$      &45.0876&map\\ 
+$^{13}$CH$_3$OH& 235997.230&$5_3-4_3$                  & 50.531 &\\
CH$_3$CHO     & 242106.023 & $13_{-1,13}-12_{-1,12}E$   & 50.2302 &rd\\ 
CH$_3$CHO     & 242118.143 & $13_{1,13}-12_{1,12}A^{++}$   & 50.1792 &rd\\ 
CH$_3$CHO     & 243178.657 & $13_{1,13}-12_{1,12}E~\nu_t=1$  & 192.2766 &rd\\ 
CH$_3$CHO     & 243975.379 & $14_{3,12}-14_{2,13}A^{+-}$   & 73.5938 &rd\\ 
CH$_3$CHO     & 244789.251 & $13_{0,13}-12_{0,12}E$     &49.6197 &map,rd\\
CH$_3$CHO     & 244832.176 & $13_{0,13}-12_{0,12}A$     &49.5596 &map,rd\\
CH$_3$CN      & 220475.822 & $12_{8}-11_{8}$ & 357.9389 &rd \\
CH$_3$CN      & 220539.335 & $12_{7}-11_{7}$ & 283.6085 &rd \\
CH$_3$CN      & 220641.089 & $12_{5}-11_{5}$ & 164.5919 &rd \\
CH$_3$CN      & 220679.287 & $12_{4}-11_4$              &119.9327&map, rd \\
CH$_3$CN      & 220709.017 & $12_{3}-11_3$              & 85.1873&map, rd \\
CH$_3$CN      & 220730.261 & $12_2-11_2$                & 60.3634&map, rd \\
CH$_3$CN      & 220743.011 & $12_1-11_1$                & 45.4669&map \\
CH$_3$CN      & 220747.262 & $12_0-11_0$                & 40.5010&map \\
CH$_3$CN      & 238843.942 & $13_{8}-12_{8}$ & 365.2932 &rd \\
CH$_3$CN      & 238912.727 & $13_{7}-12_{7}$ & 290.9649 &rd \\
CH$_3$CN      & 238972.398 & $13_{6}-12_{6}$ & 226.5128 &rd \\
CH$_3$CN      & 239022.929 & $13_{5}-12_{5}$ & 171.9517 &rd \\
CH$_3$CN      & 239064.299 & $13_4-12_4$                &127.2938&map, rd \\
CH$_3$CN      & 239096.497 & $13_3-12_3$                & 92.5493&map, rd \\
CH$_3$CN      & 239119.505 & $13_2-12_2$                & 67.7262&map, rd \\
CH$_3$CN      & 239137.917 & $13_0-12_0$                & 47.8643&map \\
CH$_3$OCH$_3$ & 222433.653 & $4_{3,1}-3_{2,2} EE$ & 7.7030 &rd \\
+CH$_3$OCH$_3$ & 222433.931 & $4_{3,1}-3_{2,2} AA$ & 7.7028 &rd \\
+CH$_3$OCH$_3$ & 222435.123 & $4_{3,1}-3_{2,2} EA$ & 7.7032 &rd \\
CH$_3$OCH$_3$ & 223200.063 & $8_{2,7}-7_{1,6} EA$ & 19.1811 &rd \\
+CH$_3$OCH$_3$ & 223200.072 & $8_{2,7}-7_{1,6} AE$ & 19.1811 &rd \\
+CH$_3$OCH$_3$ & 223202.243 & $8_{2,7}-7_{1,6} EE$ & 19.1809 &rd \\
+CH$_3$OCH$_3$ & 223204.418 & $8_{2,7}-7_{1,6} AA$ & 19.1806 &rd \\
CH$_3$OCH$_3$ & 223406.886 & $26_{2,24}-26_{1,25} AE+EA$ & 222.1807&rd \\
+CH$_3$OCH$_3$ & 223409.478 & $26_{2,24}-26_{1,25} EE$ & 222.1807 &rd \\
+CH$_3$OCH$_3$ & 223412.069 & $26_{2,24}-26_{1,25} AA$ & 222.1806 &rd \\
CH$_3$OCH$_3$ & 225598.770 & $12_{1,12}-11_{0,11} EA+AE$ & 40.9815 &rd \\
+CH$_3$OCH$_3$ & 225599.120 & $12_{1,12}-11_{0,11} EE$ & 40.9813 &rd \\
+CH$_3$OCH$_3$ & 225599.469 & $12_{1,12}-11_{0,11} AA$ & 40.9810 &rd \\
CH$_3$OCH$_3$ & 226346.124 & $14_{1,13}-13_{2,12} AA$ & 61.1644 &rd \\
+CH$_3$OCH$_3$ & 226346.948 & $14_{1,13}-13_{2,12} EE$ & 61.1646 &rd \\
+CH$_3$OCH$_3$ & 226347.772 & $14_{1,13}-13_{2,12} AE+EA$ & 61.1647 &rd \\
CH$_3$OCH$_3$ & 230140.140 & $25_{4,22}-25_{3,23} AE+EA$ & 214.1889 &rd \\
+CH$_3$OCH$_3$ & 230141.425 & $25_{4,22}-25_{3,23} EE$ & 214.1889 &rd \\
+CH$_3$OCH$_3$ & 230142.710 & $25_{4,22}-25_{3,23} AA$ & 214.1889 &rd \\
CH$_3$OCH$_3$ & 230232.166 & $17_{2,15}-16_{3,14} AA$ & 94.9454 &rd \\
+CH$_3$OCH$_3$ & 230233.749 & $17_{2,15}-16_{3,14} EE$ & 94.9454 &rd \\
+CH$_3$OCH$_3$ & 230235.333 & $17_{2,15}-16_{3,14} AE+EA$ & 94.9455 &rd \\
CH$_3$OCH$_3$ & 231987.779 & $13_{0,13}-12_{1,12} AA$ & 48.5062 &rd \\
+CH$_3$OCH$_3$ & 231987.856 & $13_{0,13}-12_{1,12} EE$ & 48.5064 &rd \\
+CH$_3$OCH$_3$ & 231987.933 & $13_{0,13}-12_{1,12} AE+EA$ & 48.5067 &rd \\
CH$_3$OCH$_3$ & 237618.821 & $9_{2,8}-8_{1,7} EA$ & 24.3654 &rd \\
+CH$_3$OCH$_3$ & 237618.826 & $9_{2,8}-8_{1,7} AE$ & 24.3654 &rd \\
+CH$_3$OCH$_3$ & 237620.888 & $9_{2,8}-8_{1,7} EE$ & 24.3651 &rd \\
+CH$_3$OCH$_3$ & 237622.953 & $9_{2,8}-8_{1,7} AA$ & 24.3649 &rd \\
CH$_3$OCH$_3$ & 240978.250 & $5_{3,3}-4_{2,2} EA$         &10.2472 &map \\
+CH$_3$OCH$_3$ & 240982.770 & $5_{3,3}-4_{2,2}AE$         &10.2472 &map, rd \\
+CH$_3$OCH$_3$ & 240985.067 & $5_{3,3}-4_{2,2} EE$         &10.2472 &map, rd \\
+CH$_3$OCH$_3$ & 240989.973 & $5_{3,3}-4_{2,2} AA$ & 10.2468 &rd \\
CH$_3$OCH$_3$ & 241528.320 & $5_{3,2}-4_{2,3} EA$ & 10.2298 &rd \\
+CH$_3$OCH$_3$ & 241528.710 & $5_{3,2}-4_{2,3} EE$ & 10.2296 &rd \\
+CH$_3$OCH$_3$ & 241531.009 & $5_{3,2}-4_{2,3} AA$ & 10.2294 &rd \\
CH$_3$OCH$_3$ & 241946.2015& $13_{1,13}-12_{0,12}EA$    &48.3184 &map, rd \\
+CH$_3$OCH$_3$& 241946.2018& $13_{1,13}-12_{0,12}AE$    &48.3184&map, rd \\
+CH$_3$OCH$_3$& 241946.4965& $13_{1,13}-12_{0,12}EE$    &48.3184&map, rd \\
+CH$_3$OCH$_3$& 241946.7913& $13_{1,13}-12_{0,12}AA$    &48.3184&map, rd \\
CH$_3$OCH$_3$ & 243738.713 & $23_{5,18}-23_{4,19} AE+EA$ & 191.3316 &rd \\
+CH$_3$OCH$_3$ & 243739.900 & $23_{5,18}-23_{4,19} EE$ & 191.3316 &rd \\
+CH$_3$OCH$_3$ & 243741.087 & $23_{5,18}-23_{4,19} AA$ & 191.3317 &rd \\
CH$_3$OCH$_3$ & 244508.305 & $23_{2,22}-23_{1,23} EE$ & 167.9316 &rd \\
+CH$_3$OCH$_3$ & 244512.709 & $23_{2,22}-23_{1,23} AA$ & 167.9314 &rd \\
CH$_3$OCHO    & 214782.311 & $18_{3,16}-17_{3,15}E$    &66.4186&rd\\
CH$_3$OCHO    & 216109.780 & $19_{2,18}-18_{2,17}E$    &68.7869&map\\
+DCO$^+$      & 216112.580 & $3_{0,0}-2_{0,0}$         & 7.2089&\\
CH$_3$OCHO    & 216966.246 & $20-19$                   & 70.2584&$^1,map$\\
CH$_3$OCHO    & 220166.809 & $17_{4,13}-16_{4,12}E$    & 64.3503&rd\\
CH$_3$OCHO    & 221693.077 & $10_{4,6}-9_{3,7}E$       & 22.6450&rd\\
CH$_3$OCHO    & 222421.356 & $18_{8,10}-17_{8,9}E$     & 92.3441&rd\\
CH$_3$OCHO    & 222657.320 & $8_{5,3}-7_{4,3}E$      & 18.8854&rd\\ 
CH$_3$OCHO    & 223162.722 & $18_{7,11}-17_{7,10}A$    & 85.4975 &rd\\ 
CH$_3$OCHO    & 225855.505 & $6_{6,1}-5_{5,1}E$        & 17.6920&rd\\  
CH$_3$OCHO    & 227019.516 & $19_{2,17}-18_{2,16}E$    & 73.4502 &rd\\
CH$_3$OCHO    & 227563.734 & $21-20$                   & 77.4957&$^1$,map\\
CH$_3$OCHO    & 228628.792 & $18_{5,13}-17_{5,12}E$    & 74.9364 &rd\\
CH$_3$OCHO    & 229405.001 & $18_{3,15}-17_{3,14}E$    & 69.3158&rd\\
CH$_3$OCHO    & 229420.343 & $18_{3,15}-17_{3,14}A$    & 69.3100&rd\\
CH$_3$OCHO    & 233226.782 & $19_{4,16}-18_{4,15}A$    & 77.8820 &rd\\
CH$_3$OCHO    & 233845.126 & $19_{11,8}-18_{11,7}E$    &125.9196 &rd\\
CH$_3$OCHO    & 235029.886 & $19_{8,11}-18_{8,10}E$    & 99.7633 &rd\\
CH$_3$OCHO    & 235263.315 & $9_{9,0}-9_{8,1}A$        & 47.9776 &rd\\ 
CH$_3$OCHO    & 235844.542 & $19_{7,13}-18_{7,12}A$    & 92.9388&rd\\
CH$_3$OCHO    & 235865.878 & $19_{7,13}-18_{7,12}E$    & 92.9388&rd\\
CH$_3$OCHO    & 236365.562 & $20_{3,18}-19_{3,17}A$    & 81.1013&rd\\
CH$_3$OCHO    & 236743.645 & $19_{5,15}-18_{5,14}E$    & 82.1900&rd\\
CH$_3$OCHO    & 236800.513 & $19_{6,14}-18_{6,13}E$    & 87.1016&rd\\
CH$_3$OCHO    & 236810.327 & $19_{6,14}-18_{6,13}A$    & 87.0979&rd\\
CH$_3$OCHO    & 237398.615 & $21_{2,20}-20_{1,19}A$    & 83.5457&rd\\
CH$_3$OCHO    & 237807.577 & $19_{6,13}-18_{6,12}E$    & 87.1428&rd\\
CH$_3$OCHO    & 237829.829 & $19_{6,13}-18_{6,12}A$    & 87.1385&rd\\
CH$_3$OCHO    & 238157.297 & $22-21$                   & 85.0863&$^1$,map\\ 
CH$_3$OCHO    & 240021.129 & $19_{3,16}-18_{3,15}E$    & 76.9679&rd\\
CH$_3$OCHO    & 242871.513 & $19_{5,14}-18_{5,13}E$    & 82.5627&rd\\
CH$_3$OCHO    & 242896.022 & $19_{5,14}-18_{5,13}A$    & 82.5584&rd\\
CH$_3$OCHO    & 244580.313 & $20_{4,17}-19_{4,16}E$    & 85.6670&rd\\
CH$_3$OCHO    & 246891.590 & $19_{4,15}-18_{4,14}E$    & 79.4914&rd\\
CH$_3$OCHO    & 249047.435 & $20_{5,16}-19_{5,15}A$    & 90.0826&rd\\
CH$_3$OH      & 211803.245 & $ 16_{2,15}-15_{1,14} A^- \nu_t=1 $ & 419.2430 & rd \\
CH$_3$OH      & 213159.369 & $ 20_{-4,17}-19_{-5,14} E $ & 392.4940 & rd \\
CH$_3$OH      & 213377.521 & $ 13_{6,8}-14_{5,10} E $ & 263.8890 & rd \\
CH$_3$OH      & 213427.061 & $1_{1}-0_{0}E$& 9.1220 &map, rd \\
CH$_3$OH      & 215302.205 & $ 6_{1,6}-7_{2,5} A^+ \nu_t=1$ & 252.6430 & rd \\
CH$_3$OH      & 216945.559 & $ 5_{1,4}-4_{2,2} E $ & 31.5960 & rd \\
CH$_3$OH      & 217299.205 &$6_{1}-7_{2}A^{-} \nu_t=1$ & 252.6437&map, rd \\
CH$_3$OH      & 217886.39 & $ 20_{1,19}-20_{0,20} E $ & 346.0730 & rd \\
CH$_3$OH      & 218440.063 &$4_{2}-3_{1}E$&24.3097&map, rd \\
CH$_3$OH      & 220078.561 & $8_{0}-7_{1}E$& 59.8092&map, rd \\
CH$_3$OH      & 222722.796 & $ 16_{2,14}-15_{1,15} A^+ \nu_t=1 $ & 418.8610 & rd \\
CH$_3$OH      & 224699.714 & $ 20_{-2,19}-19_{-3,17} E $ & 349.9350 & rd \\
CH$_3$OH      & 227094.600  & $ 21_{1,20}-21_{0,21} E $ & 379.6110 & rd \\
CH$_3$OH      & 227814.528 & $16_{1}-15_{2}A^{+}$&219.8440&map, rd \\
CH$_3$OH      & 229589.056 & $15_{4}-16_{3}E$&252.5906&map, rd \\
CH$_3$OH      & 229758.811 & $ 8_{-1,8}-7_{0,7} E $ & 54.2660 & rd \\
CH$_3$OH      & 229939.180 & $ 19_{5,14}-20_{4,17} A^- $ & 394.4770& rd \\
CH$_3$OH      & 230027.002 & $ 3_{-2,2}-4_{-1,4} E $ & 20.0090 & rd \\
CH$_3$OH      & 231281.110 & $10_{2}-9_{3}A^{-}$&107.2083&map, rd \\
CH$_3$OH      & 232418.571 & $ 10_{2,8}-9_{3,7}A^+ $ & 107.2080 & rd \\
CH$_3$OH      & 232783.591 & $ 18_{3,16}-17_{4,13} A^+ $ & 302.5920 & rd \\
CH$_3$OH      & 232945.797 & $10_{-3}-11_{-2}E$&124.5436&map, rd \\
CH$_3$OH      & 233795.799 & $ 18_{3,15}-17_{4,14} A^- $ & 302.5920 & rd \\
CH$_3$OH      & 234683.370 & $4_{2}-5_{1}A^{-}$&34.5156&map, rd \\ 
CH$_3$OH      & 234698.519 &$5_{-4}-6_{-3}E$&77.4675&map, rd \\
CH$_3$OH      & 236936.089 &$14_{1}-13_{2}A^{-}$&172.9481&map, rd \\
CH$_3$OH      & 237129.23 & $ 22_{1,21}-22_{0,22} E $ & 414.7270 & rd \\
CH$_3$OH      & 239746.219 &$5_{1}-4_{1}A^{+}$   &26.1012&map, rd \\
CH$_3$OH      & 240241.490 & $5_{3}-6_{2}E$       &49.3488&map, rd \\
CH$_3$OH      & 240960.557 &$5_{1}-4_{1}A^{+},~ \nu_t=1$&242.1426&map \\
CH$_3$OH      & 241206.035 & $5_{0}-4_{0}E,~ \nu_t=1$ &225.0053&map \\
CH$_3$OH      & 241238.108 & $ 5_{-1,4}-4_{-1,3} E \nu_t=1 $ & 303.4060 & rd \\
CH$_3$OH      & 241267.822 & $ 5_{0,5}-4_{0,4} A^+ \nu_t=1$ & 310.5470 & rd \\
CH$_3$OH      & 241364.12 & $ 5_{1,4}-4_{1,3} A^- \nu_t=2 $ & 490.6630 & rd \\
CH$_3$OH      & 241441.265 & $ 5_{1,4}-4_{1,3} A^- \nu_t=1 $ & 242.1750 & rd \\
CH$_3$OH      & 241700.159 & $5_{0}-4_{0}E$        & 25.2542&map \\
CH$_3$OH      & 241767.234 &$ 5_{-1}-4_{-1}E$      &20.0091 &map \\
CH$_3$OH      & 241791.352 & $5_{0}-4_{0}A^+$      & 16.1335&map, rd \\
CH$_3$OH      & 241806.525 &$ 5_{4}-4_{4}A^{+}$   &71.9750 &map \\
              &            & $+5_{4}-4_{4}A^{-}$  &        &map \\
CH$_3$OH      & 241813.255 & $5_{-4}-4_{-4}E$      & 77.2302&map \\
CH$_3$OH      & 241832.718 & $5_{3}-4_{3}A^{+}$   & 50.7463&map \\
              &            & $+5_{3}-4_{3}A^{-}$  &        &map \\
CH$_3$OH      & 241843.604 & $5_{3}-4_{3}E$        & 49.2953&map \\
CH$_3$OH      & 241852.299 & $5_{-3}-4_{-3}E$      & 59.7198&map \\
CH$_3$OH      & 241879.025 & $5_{1}-4_{1}E$        & 30.7644&map \\
CH$_3$OH      & 241887.674 & $5_{2}-4_{2}A^{+}$   & 42.3450&map \\
CH$_3$OH      & 241904.147 & $5_{2}-4_{2}E$        & 34.1368&map \\
+CH$_3$OH     & 241904.643 & $5_{-2}-4_{-2}E$      & 31.5961&map \\
CH$_3$OH      & 242446.084 & $14_{-1}-13_{-2}E$  &164.9296&map, rd \\
CH$_3$OH      & 243413.43 & $ 23_{3,20}-23_{2,21} A^{-+}$ & 471.5150 & rd \\
CH$_3$OH      & 243915.788 & $5_{1}-4_{1}A^{-}$   & 26.3795&map, rd \\
CH$_3$OH      & 244330.987 & $ 22_{3,19}-22_{2,20} A^{-+}$ & 434.4160 & rd \\
CH$_3$OH      & 244338.004 & $ 9_{1,9}-8_{0,8} E \nu_t=1$ & 266.8380 & rd \\
CH$_3$OH      & 245223.465 & $ 21_{3,18}-22_{2,19} A^{-+} $ & 398.9280 & rd \\
CH$_3$OH      & 246074.914 & $ 20_{3,17}-20_{2,18} A^{-+} $ & 365.0510 & rd \\
CH$_3$OH      & 246873.503 & $ 19_{3,16}-19_{2,17} A^{-+} $ & 332.7850 & rd \\
CH$_3$OH      & 247161.950 & $16_{2}-15_{3}E$    &226.7747&map, rd \\
CH$_3$OH      & 247228.587 & $4_{2}-5_{1} A^{+}$  & 34.0983&map, rd \\
CH$_3$OH      & 247611.037 & $ 18_{3,15}-18_{2,16} A^{-+} $ & 302.1310 & rd \\
CH$_3$OH      & 247840.224 & $ 12_{-2,10}-13_{-3,10} E \nu_t=1 $ & 370.6090 & rd \\
CH$_3$OH      & 248282.424 & $17_{3}-17_{2}A^{-+}$&273.0892&map \\
CH$_3$OH      & 248885.468 & $16_{3}-16_{2}A^{-+}$&245.6602&map, rd \\
CH$_3$OH      & 249192.864 & $ 16_{-3,14}-15_{-4,12} E $ & 254.6040 & rd \\
CH$_3$OH      & 249419.924 & $15_{3}-15_{2}A^{-+}$&219.8440&map, rd \\
CH$_3$OH      & 249443.402 & $ 7_{4,4}-8_{3,5} A^- $ & 92.6910 & rd \\
CH$_3$OH      & 249451.885 & $ 7_{4,3}-8_{3,6} A^+ $ & 92.6900 & rd \\
CN            &    226359.987  & $2-1~J=3/2-3/2$            &3.7857  &map \\
              &            & $~~~~F=5/2-5/2$            &        & \\
CN            & 226632.190 & $2-1~J=3/2-1/2$            & 3.7757 &map \\
              &            & $~~~~F=3/2-3/2$            &        & \\
CN            & 226659.575 & $2-1~J=3/2-1/2$            & 3.7757 &map \\
              &            & $~~~~F=5/2-3/2$            &        & \\
CN            & 226679.382 & $2-1~J=3/2-1/2$            & 3.7741 &map \\
              &            & $~~~~F=3/2-1/2$            &        & \\
CN            & 226875.897 & $2-1~J=5/2-3/2$            & 3.7866 &map \\
              &            & $~~~~F=3/2-1/2$            &        & \\
CS            & 244935.644 & $5-4 $ & 16.3411           &map \\
H$_2^{13}$CO  & 219908.525 & $3_{1,2}-2_{1,1}$          & 15.5579&map \\
H$_2$CO       & 211211.468 & $ 3_{1,3}-2_{1,2}$         & 15.2369&map, rd \\
H$_2$CO       & 216568.668 & $9_{1,8}-9_{1,9}$ & 113.7050 &rd \\
H$_2$CO       & 218222.192 & $3_{0,3}-2_{0,2}$          & 7.2864 &map, rd \\
H$_2$CO       & 218475.632 & $3_{2,2}-2_{2,1}$          &40.0402 &map, rd \\
H$_2$CO       & 218760.066 & $3_{2,1}-2_{2,0}$          &40.0426 &map, rd \\
H$_2$CO       & 225697.775 & $3_{1,2}-2_{1,1}$          &15.7202 &map, rd \\
H$_2$CS       & 236726.770 & $7_{1,7}-6_{1,6}$          &32.8599 &map \\
H$_2$CS       & 240266.320 & $7_{0,7}-6_{0,6}$          &24.0546 &map \\
H$_2$CS       & 244047.840 & $7_{1,6}-6_{1,5}$          &33.5928 &map \\
H$_2$S        & 216710.435 & $2_{2,0}-2_{1,1}$          &51.1402 &map \\
HCCCN         & 236512.777 & $26-25$ & 98.6235          &map \\
HCCCN         & 245606.308 & $27-26$ & 106.5127         &map \\
HCOOH         & 220037.960 & $10_{0,10}-9_{0,9}$        &33.4039  &rd \\
HCOOH         & 223915.560 & $10_{2,9}-9_{2,8}$         &42.5246  &rd \\
HCOOH         & 225085.440 & $10_{4,7}-9_{4,6}$         &69.1174  &rd \\
HCOOH         & 225091.210 & $10_{4,6}-9_{4,5}$         &69.1175  &rd \\
HCOOH         & 225512.540 & $10_{3,7}-9_{3,6}$         &53.6767  &rd \\
HCOOH         & 228544.070 & $10_{2,8}-9_{2,7}$         &42.8481  &rd \\
HCOOH         & 231505.590 & $10_{1,9}-9_{1,8}$         &37.0844  &rd \\
HCOOH         & 236717.326 & $11_{1,11}-10_{1,10}$      &41.8215 &map, rd \\
HCOOH         & 241146.200 & $11_{0,11}-10_{0,10}$      &40.7435  &rd \\
HCOOH         & 246106.087 & $11_{2,10}-10_{2,9}$       &49.9937 &map, rd \\
HCOOH         & 247514.120 & $11_{*,6}-10_{*,5}$        &96.4903  &rd \\
HCS$^+$       & 213360.530 & $5-4 $ & 14.2343           &map \\
HNCO          & 218981.031 & $10_{1,10}-9_{1,9}$        &62.9493 &map, rd \\
HNCO          & 219656.805 & $10_{3,7}-9_{3,6} $        &303.6759 & rd \\
              &            & $+10_{3,8}-9_{3,7}$        &         & rd \\
HNCO          & 219733.824 & $10_{2,9}-9_{2,8}$         &153.2967 &rd \\
HNCO          & 219798.247 & $10_{0,10}-9_{0,9}$        &32.9940 &map, rd \\
HNCO          & 220585.200 & $10_{1,9}-9_{1,8}$         &63.1900 &map, rd \\
HNCO          & 231873.255 & $28_{1,28}-29_{0,29}$      &318.8428 & rd \\
HNCO          & 240875.856 & $11_{1,11}-10_{1,10}$      &70.2537 &map, rd \\
HNCO          & 241774.037 & $11_{0,11}-10_{0,10}$      &40.3257 & rd \\
HNCO          & 242639.705 & $11_{0,10}-10_{0,9}$       &70.5480 & rd \\
OCS           & 231060.983 & $19-18 $ & 69.3719         &map \\
OCS           & 243218.040 & $20-19  $                  &77.0793 &map \\
SO            & 215220.653 & $N,J=5,5-4,4 $             &23.4748 &map \\
SO            & 219949.442 & $N,J=5,6-4,5 $             &16.9790 &map \\
$^{34}$SO       & 215839.920 & $N,J=5,6-4,5$            & 16.6981&map \\
SO$_2$        & 221965.210 & $11_{1,11}-10_{0,10}$      &34.5495 &map \\
SO$_2$        & 224264.811 & $20_{2,16}-19_{3,17}$      &136.9168&map \\
SO$_2$        & 225153.702 & $13_{2,12}-13_{1,13}$      &57.1170 &map \\
SO$_2$        & 226300.027 & $14_{3,11}-14_{2,12}$      &75.1496 &map \\
SO$_2$        & 235151.720 & $4_{2,2}-3_{1,3}$          & 5.3820 &map \\
SO$_2$        & 236216.685 & $16_{1,15}-15_{2,14}$      &82.9373 &map \\
SO$_2$        & 237068.870 & $12_{3,9}-12_{2,10}$       &57.3978 &map \\
SO$_2$        & 241615.798 & $5_{2,4}-4_{1,3}$          & 8.3356 &map \\
SO$_2$        & 244254.218 & $14_{0,14}-13_{1,13}$      &57.1170 &map \\
SO$_2$        & 245563.423 & $10_{3,7}-10_{2,8}$        &42.3467 &map \\
\hline
\end{longtable}
$^1$--blend of several CH$_3$OCHO transitions.

\end{document}